\documentclass[12pt,letterpaper]{article}
\pdfoutput=1
\usepackage[pdftex]{graphicx,color}
\usepackage{hyperref}
\input{psfig}
\usepackage{amsmath,amssymb}
\usepackage[dvips,letterpaper,text={6.5in,9in}]{geometry}
\usepackage{fancyhdr}
\usepackage{verbatim}

\usepackage{dcolumn}   % needed for some tables 
\usepackage{subfig}
\usepackage{float}
\usepackage{ytableau}
\newcommand\ltap{\
  \raise.3ex\hbox{$<$\kern-.75em\lower1ex\hbox{$\sim$}}\ }
\newcommand\gtap{\
  \raise.3ex\hbox{$>$\kern-.75em\lower1ex\hbox{$\sim$}}\ }
\newcommand\simge{\mathrel{%
   \rlap{\raise 0.511ex \hbox{$>$}}{\lower 0.511ex \hbox{$\sim$}}}}
\newcommand\simle{\mathrel{
   \rlap{\raise 0.511ex \hbox{$<$}}{\lower 0.511ex \hbox{$\sim$}}}}

\newcommand{\slashchar}[1]%
        {\kern .25em\raise.18ex\hbox{$/$}\kern-.75em #1}
\def\lsim{\mathrel{\raise.3ex\hbox{$<$\kern-.75em\lower1ex\hbox{$\sim$}}}}
\def\gsim{\mathrel{\raise.3ex\hbox{$>$\kern-.75em\lower1ex\hbox{$\sim$}}}}
\newcommand{\bs}{\boldsymbol}

\newcommand\CH{{\cal H}}

\newcommand\CL{{\cal L}}

\newcommand\CO{{\cal O}}

\newcommand\CQ{{\cal Q}}

\newcommand\be{\begin{equation}}
\newcommand\ee{\end{equation}}
\newcommand\bea{\begin{eqnarray}}
\newcommand\eea{\end{eqnarray}}
\newcommand\ba{\begin{array}}
\newcommand\ea{\end{array}}
\newcommand\nn{\nonumber}

\newcommand{\half}{\ensuremath{\frac{1}{2}}}

\newcommand{\thalf}{\textstyle{\frac{1}{2}}}
\newcommand{\tthalf}{\textstyle{\frac{3}{2}}}

\newcommand{\thw}{\ensuremath{\theta_W}}

\newcommand\ra{\rightarrow}

\newcommand\gev{{\rm GeV}}
\newcommand\tev{{\rm TeV}}

\newcommand\pb{{\rm pb}}

\newcommand\fb{{\rm fb}}
\newcommand\ifb{{\rm fb}^{-1}}

\newcommand\lvac{\langle \Omega \vert}
\newcommand\rvac{\vert \Omega \rangle}

\newcommand\ellm{\ell^-}
\newcommand\ellpm{\ell^\pm}
\newcommand\ellp{\ell^+}

\newcommand\getc{g_{ETC}}

\newcommand\Ntc{N_{TC}}

\newcommand\tom{\omega_{T}}
\newcommand\tro{\rho_{T}}

\newcommand\trho{\rho_T}
\newcommand\ta{a_T}

\newcommand\tropm{\rho_{T}^\pm}

\newcommand\troz{\rho_{T}^0}

\newcommand\tpi{\pi_T}
\newcommand\tpipm{\pi_T^\pm}
\newcommand\tpimp{\pi_T^\mp}

\newcommand\tpiz{\pi_T^0}
\newcommand\tpipr{\pi_T^{0 \prime}}

\newcommand\etat{\eta_{_{T}}}
\newcommand\etal{\eta_{_{L}}}
\newcommand\etah{\eta_{_{H}}}
\newcommand\Fetat{F_{\eta_T}}

\newcommand\Mjj{M_{jj}}

\newcommand{\antisymm}{\ytableausetup{boxsize=0.65em}\begin{ytableau} \\
    \\ \end{ytableau} }
\newcommand{\fund}{\ytableausetup{boxsize=0.65em}\begin{ytableau}
    \\ \end{ytableau}}

\begin{document}

\title{
\vskip -15mm
\begin{flushright}
 \vskip -15mm
 {\small FERMILAB-Pub-12-573-T\\
   CERN-PH-TH/2012-273\\
 }
 \vskip 5mm
 \end{flushright}
{\Large{\bf A Higgs Impostor in Low-Scale Technicolor }}\\
} \author{
  {\large Estia Eichten$^{1}$\thanks{eichten@fnal.gov}},\,
  Kenneth Lane$^{2}$\thanks{lane@physics.bu.edu},\,
\, and Adam Martin$^{3,4}$\thanks{adam.martin@cern.ch}\\
{\large {$^{1}$}Theoretical Physics Group, Fermi National Accelerator
  Laboratory}\\
{\large P.O. Box 500, Batavia, Illinois 60510}\\
{\large $^{2}$Department of Physics, Boston University}\\
{\large 590 Commonwealth Avenue, Boston, Massachusetts 02215}\\
{\large $^{3}$PH-TH Department, CERN}\\
{\large CH-1211 Geneva 23, Switzerland} \\
{\large $^{4}$Department of Physics, University of Notre
  Dame\footnote{Visiting scholar}}\\
{\large Notre Dame. Indiana 46556} \\
}
\maketitle

\begin{abstract}
  
  We propose a ``Higgs impostor'' model for the 125~GeV boson, $X$, recently
  discovered at the LHC. It is a technipion, $\etat$, with $I^G J^{PC} = 0^-
  0^{-+}$ expected in this mass region in low-scale technicolor. Its coupling
  to pairs of standard-model gauge bosons are dimension-five operators whose
  strengths are determined within the model. It is easy for the gluon fusion
  rate $\sigma B(gg \to \etat \to \gamma\gamma)$ to agree with the measured
  one, but $\etat \to ZZ^*,\,WW^*$ are greatly suppressed relative to the
  standard-model Higgs rates. This is a crucial test of our proposal. In this
  regard, we assess the most recent data on $X$~decay modes, with a critical
  discussion of $X \to ZZ^* \to 4\ell$. In our model the $\etat$ mixes almost
  completely with the isovector $\tpiz$, giving two similar states, $\etal$
  at $125\,\gev$ and $\etah$ higher, possibly in the range 170--$190\,\gev$.
  Important consequences of this mixing are (1)~the only associated
  production of $\etal$ is via $\tro \to W \etal$, and this could be sizable;
  (2)~$\etah$ may soon be accessible in $gg \to \etah \to \gamma\gamma$; and
  (3)~LSTC phenomenology at the LHC is substantially modified.

\end{abstract}

%%%%%%%%%%%%%%%%%%%%%%%%%%%%%%%%%
%%%%%%%%%%%%%%%%%%%%%%%%%%%%%%%%%

\newpage

\section*{1. Introduction}

The stunning discovery by ATLAS and CMS of a new boson $X$ at $125\,\gev$
decaying into $\gamma\gamma$ and, at lower significance, $ZZ^*$ and
$WW^*$~\cite{:2012gk,:2012gu} is widely suspected to be the long-sought Higgs
boson of the standard model (SM) of electroweak
interactions~\cite{Glashow:1961tr,Weinberg:1967tq,Salam:1968rm,Englert:1964et,
  Higgs:1964ia,Guralnik:1964eu}. It is also widely believed that the
collaborations' latest releases of data~\cite{CMSPASHIG,CMStwiki,
  ATLASCONFggZZ,ATLASCONFZZ} strongly support this
suspicion~\cite{PausHCP,EinsweilerHCP}. However, as emphasized by Wilson
(quoted in Ref.~\cite{Susskind:1978ms}) and `t~Hooft~\cite{'tHooft:1979bh}
this explanation for the origin of electroweak symmetry breaking is very
unsatisfactory. It is beset by the great problems of naturalness, hierarchy
and flavor---the number, masses and mixings of the fermion generations.
Notwithstanding this, the discovery clearly puts great pressure on
technicolor, the scenario for the Higgs mechanism which needs no Higgs-like
boson~\cite{Weinberg:1979bn,Susskind:1978ms}. This is especially true in
low-scale technicolor (LSTC)~\cite{Lane:1989ej, Lane:2009ct}. As far as we
understand, there is no LSTC bound state that mimics $H$-decays in all these
channels and at the rates expected on the basis of the observed $\sigma(pp
\to X)B(X \to \gamma\gamma)$.\footnote{There is a low-lying $I^G J^{PC} = 0^+
  0^{++}$ state in LSTC with many of the same decays as the standard model
  $H$, but its production rate is too small to be the boson observed at the
  LHC~\cite{Delgado:2010bb}}$^,$\footnote{It is argued by some that walking
  technicolor or similar models have a light scalar due to their
  near-conformal invariance being spontaneously broken. This is called the
  ``techni-dilaton''. It is also argued that it has Higgs-like couplings to
  gauge bosons and fermions; see e.g., Refs.~\cite{Cheung:2011nv,
    Matsuzaki:2012gd, Matsuzaki:2012fq,Dietrich:2005jn,Elander:2012fk,
    Chacko:2012vm, Bellazzini:2012vz}. In our view, the existence of such a
  state is questionable. An interesting paper that discusses the
  phenomenology of a light dilaton while merely assuming its existence is
  Ref.~\cite{Goldberger:2007zk}.}

In this paper we propose that $X(125)$ is a state expected in a two-scale
model of LSTC and which may be consistent with the data made public so far.
This state is a would-be axion, a mixture of neutral isoscalar pseudoscalars
occurring in each scale-sector that would be nearly massless if it were not
for extended technicolor (ETC) interactions connecting the technifermions of
the two scales. We call this particle the $\etat$. It has $CP =
-1$.\footnote{In addition to the dilaton papers cited above, others that have
  recently suggested a pseudoscalar Higgs impostor in the context of strong
  electroweak symmetry breaking include Refs.~\cite{Low:2011gn,
    Burdman:2011ki,Holdom:2012pw,Low:2012rj,Chivukula:2012cp, Coleppa:2012eh,
    Frandsen:2012rj}. Unlike our model, most of these do not determine the
  energy scale and other factors in the dimension-five operators that couple
  the pseudoscalar to a pair of SM gauge bosons; see
  Eqs.~(\ref{eq:etagg})--(40).} As we will see, $\sigma B(pp \to \etat \to
\gamma\gamma$) can be larger than the corresponding SM Higgs cross section,
and easily match the current experimental observation. In the model we study,
there is an unanticipated and interesting possibility: the $\etat$ mixes,
probably very substantially, with the neutral isovector $\tpiz$ expected in
LSTC. This results in two states, $\etal$ at $125\,\gev$ and a heavier state
$\etah$ which, we will argue, is likely to be at 170--$190\,\gev$. They have
similar production and decay modes, characteristic of both $\etat$ and
$\tpiz$. We shall refer to our Higgs impostor as $\etat$ in the absence of
large mixing, or as $\etal$ if mixing is important.

First, however, we ask: is $X(125)$ a Higgs boson? If analyses of the data in
hand, approximately $5\,\ifb$ at $7\,\tev$ and $20\,\ifb$ at $8\,\tev$,
establish that the rates for $pp \to X \to ZZ^*$ and $WW^*$ are in accord
with the standard model and that $X \to \tau^+\tau^-$ and $\bar bb$ are
convincingly seen at Higgsish rates, it will be difficult to resist the
conclusion that $X$ is a Higgs boson, perhaps even the SM Higgs boson, $H$.
At this early stage of $X$-physics studies, however, there are several
possibly statistical peculiarities and discrepancies with the standard model
or between the experiments~\cite{:2012gk,:2012gu,Incandela:2012in,
  CMSPASHIG,CMStwiki, ATLASCONFggZZ, ATLASCONFZZ}that allow for an
alternative explanation. These are discussed in Sec.~2 with special
attention to the high mass-resolution process $X \to ZZ^* \to 4\ell$.

In Sec.~3 we present a two-scale model for the $\etat$. This model is not
unique, but it is simple. Because the $\etat$ is a pseudo-Goldstone boson of
a chiral symmetry spontaneously broken to a vectorial one, it has $CP = -1$
and all its interactions with a pair of SM gauge bosons are of the
nonrenormalizable Wess-Zumino-Witten (WZW)
type~\cite{Wess:1971yu,Witten:1983tw}.
In Sec.~4 we discuss mixing of the isoscalar $\etat$ with the isovector
$\tpiz$. This mixing is essentially complete in our model and it may be a
general feature of two-scale models with rather widely separated energy
scales. This gives two states, $\etal$ at $125\,\gev$ and a similar state
$\etah$ at higher mass. If the dijet excess reported by
CDF~\cite{Aaltonen:2011mk} is real and is described by
LSTC~\cite{Eichten:2011sh} then we predict $M_{\etah} = 170$--$190\,\gev$. We
urge a search for such a state decaying to two
photons.
% Furthermore, this mixing
% opens up the possibility of associated production of $\etal$ with $W$, but
% {\em not} $Z$, via the resonant process $\tropm \to W^\pm \etal$.
The WZW interactions of our Higgs impostor are determined in Sec.~5 for the
unmixed and mixed cases. Compared to the SM Higgs boson, they imply very
little $\etat \to WW^*,\, ZZ^*$ and vector boson fusion (VBF) of $\etat$ via
$WW$ and $ZZ$. There is also little associated production of $\etat$ with $W$
or $Z$ {\em unless} it mixes appreciably with $\tpiz$. In that case, and
assuming the validity of the CDF $Wjj$ data, $\tropm \to \etal W^\pm$ readily
occurs, but {\em not} $\troz \to \etal Z$. Decays of $\etal$ are dominated by
$\etat \to gg$ and, so, $\etal$ decays nearly 100\% of the time to $gg$. This
may pollute and alter the SM $WW/WZ \to \ell\nu jj$ signal {\em and} the CDF
dijet excess. We also discuss $\etat$ couplings to fermion pairs; these are
induced by ETC and are, therefore, rather uncertain.

The phenomenology of $\etal$ is presented in Sec.~6. In detail, it is
specific to our two-scale model, but the general features, especially those
dictated by the WZW interactions, hold in any such model. In particular:
(1)~By far, the dominant $\etal$-production mechanism is via gluon
fusion. % Associated production with $W$ depends on the mixing of $\etat$ with
% $\tpiz$. There is very little associated production with~$Z$, and very little
% VBF via $WW$ and $ZZ$.
Generally, we find that $\sigma(gg \to \etal) > \sigma(gg \to H)$. Obtaining
the correct $\sigma B(gg \to \etal \to \gamma\gamma)$ rate is then due to a
fortuitous (but ubiquitous) cancellation among the terms in the
$\gamma\gamma$ amplitude. (2)~As noted, the branching ratios $B(\etal \to
ZZ^*,\, WW^* \to \,\, {\rm leptons})$ are extremely small.  Therefore,
according to our model's framework, what has been observed by CMS and ATLAS
must be background. The current experimental situation, which we critique in
Sec.~2, still allows this possibility. (3)~The branching ratios of $\etal$ to
$\tau^+\tau^-$ and $\bar bb$ depend on the unknown couplings of $\etat$ and
$\tpiz$ to these fermions in the underlying ETC model. We fix them to be
consistent with current data. In Sec.~7 we summarize the consequences of
$\etat$-$\tpiz$ mixing on the low-scale $\tro$ phenomenology at the LHC.
These are dramatic if the mixing is as large as we find in Sec.~4, and we
expect it to be more difficult to detect the signatures we discussed in
Ref.~\cite{Eichten:2012hs}.

\section*{2. $X$-Decay Data in 2012}

The new boson $X$ is widely referred to as being ``Higgs-like'' because it
{\em appears} to have been observed in several of the experimentally most
accessible decay channels of the SM Higgs boson, namely $\gamma\gamma$, $ZZ^*
\to \ellp\ellm\ellp\ellm$ ($\ell = e$ and/or~$\mu$), $WW^* \to
\ellp\nu\ellm\nu$, $\tau^+\tau^-$ and $WX \to \ell\nu \bar bb$. Furthermore,
$\sigma B$ for these channels are roughly consistent with those predicted for
a standard model Higgs of mass $125\,\gev$. We say ``appears'' because, as we
now discuss, the evidence for some of these decay channels is rather weak
and, we believe, the important $ZZ^* \to 4\ell$ channel is still dominated by
statistics.

\begin{figure}[!t]
 \begin{center}
\includegraphics[width=3.15in,height=3.15in]{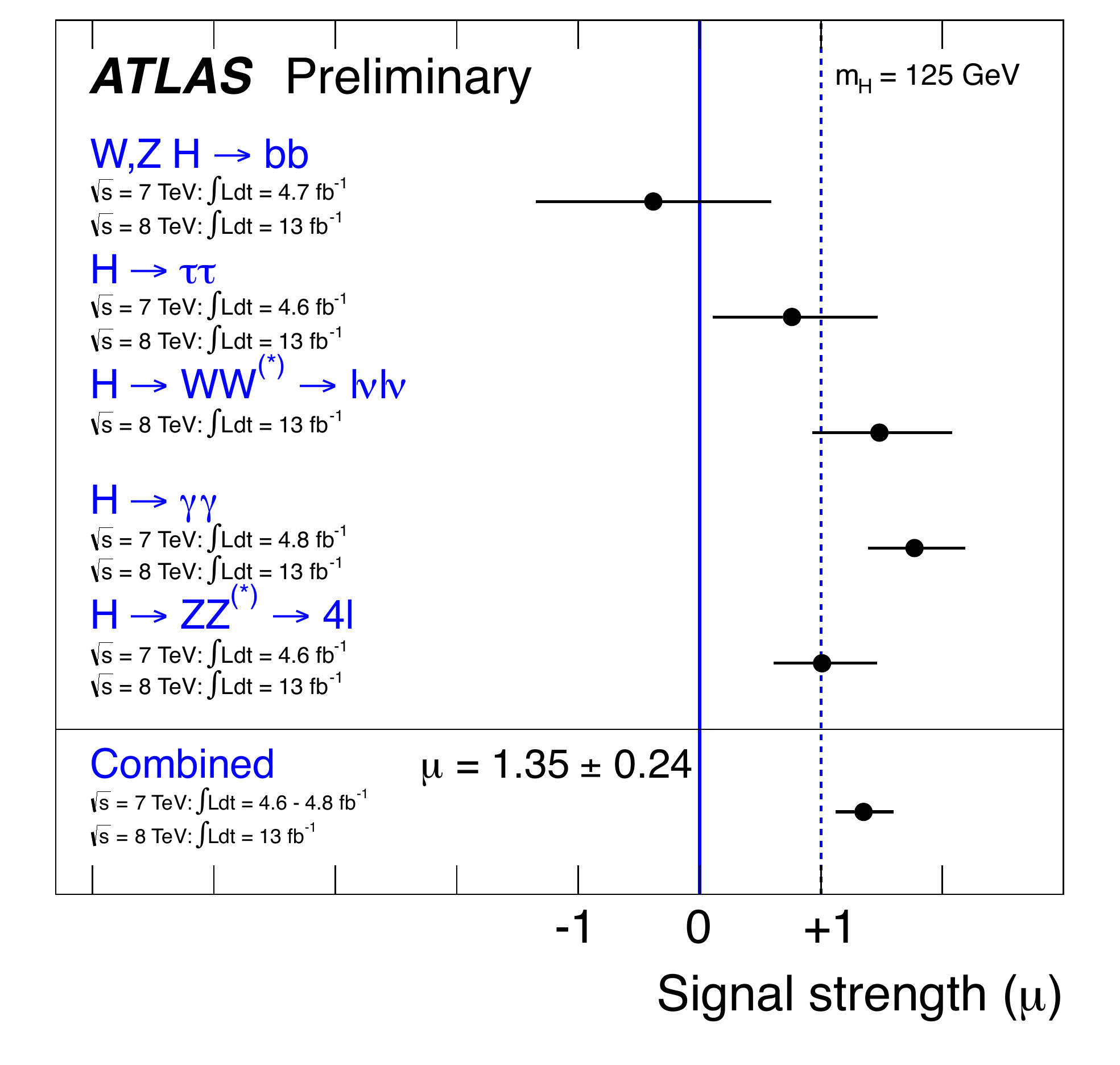}
\includegraphics[width=3.15in, height=3.15in]{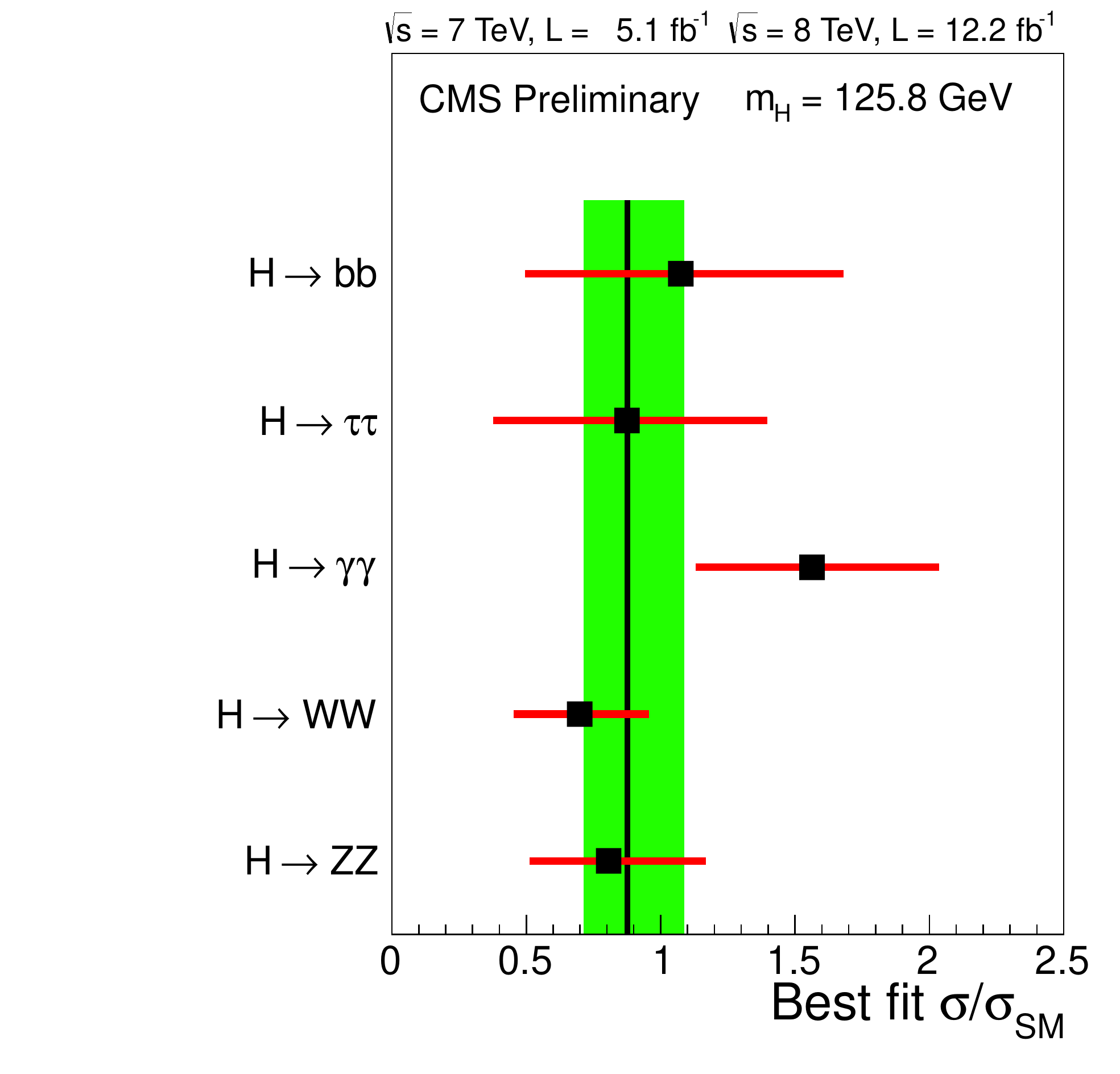}
%\vskip -0.5in
\caption{The signal strengths $\mu(F) = \sigma B(pp\to X\to F)/\sigma B(pp\to
  H\to F)$ determined by ATLAS as of December 2012~\cite{ATLASCONFggZZ}
  (left) and CMS as of November 2012~\cite{CMSPASHIG} (right) for
  luminosities of about $5\,\ifb$ at $7\,\tev$ and 12--$13\,\ifb$ at
  $8\,\tev$ (except that CMS's $\gamma\gamma$ data at $8\,\tev$ is based on
  only $5\,\ifb$).
  \label{fig:mufig}}
 \end{center}
 \end{figure}
\begin{figure}[!t]
 \begin{center}
\includegraphics[width=3.15in, height=3.00in]{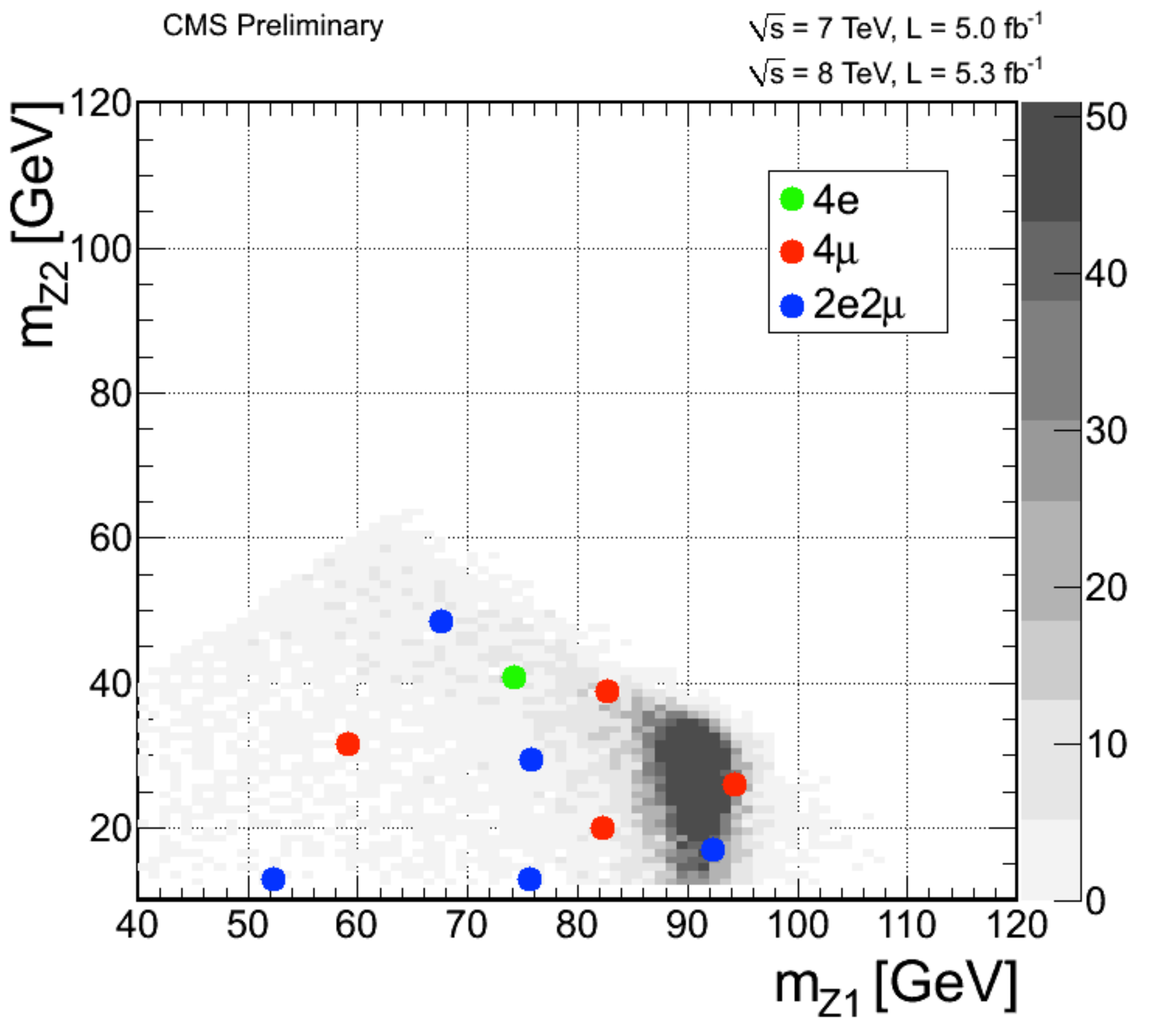}
\includegraphics[width=3.15in, height=2.90in]{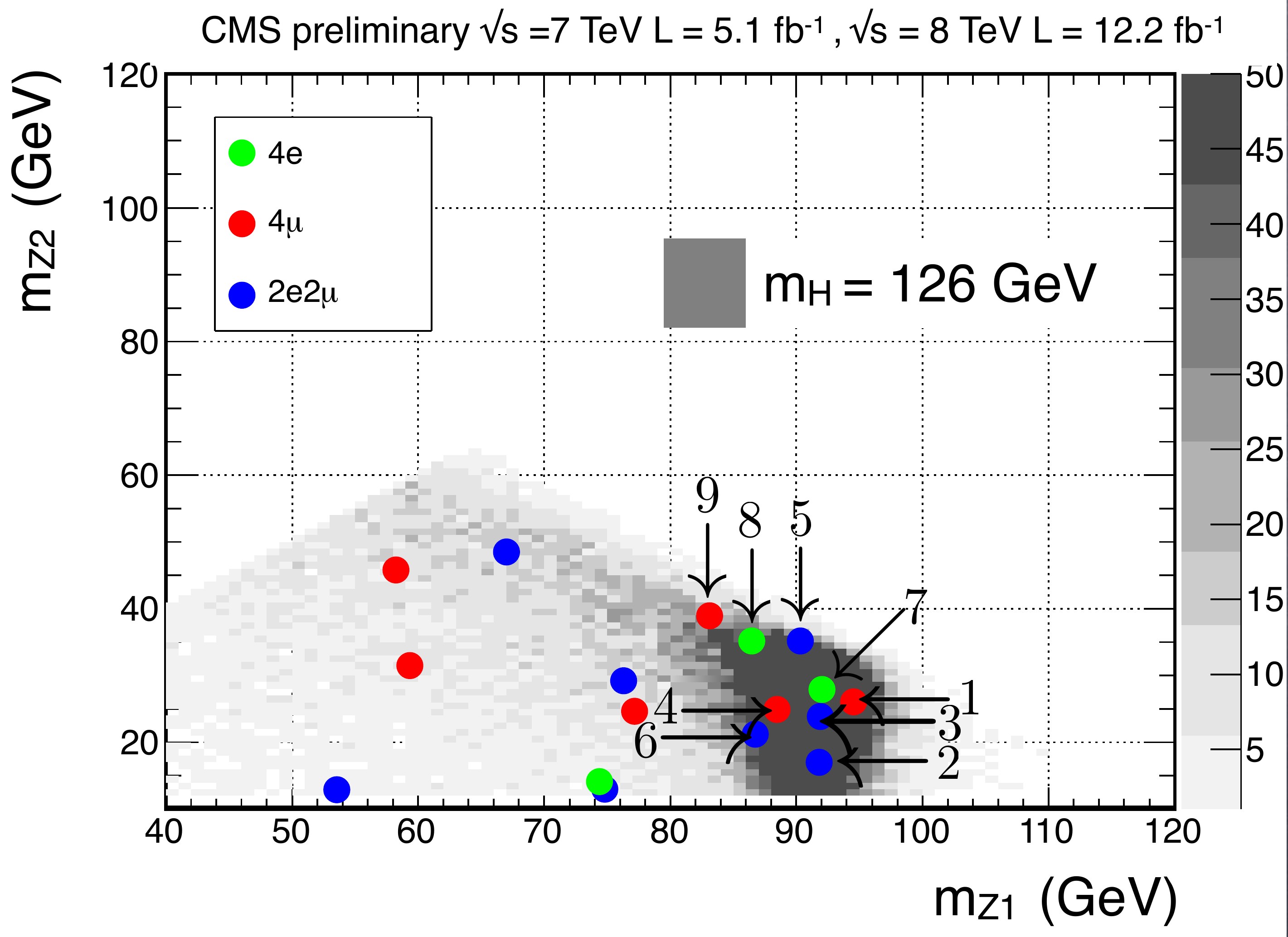}
%\vskip -0.5in
\caption{The Dalitz plot of high vs.~low dilepton mass, $M_{Z1}$ vs.~$M_{Z2}$
  in the four-lepton invariant mass region $120\,\gev < M_{4\ell} <
  130\,\gev$ from CMS in July at ICHEP-2012~\cite{Incandela:2012in} (left)
  and November 2012~\cite{CMSHigtwiki} (right). We have numbered
  ``signal-like'' events as described in the text.
  \label{fig:CMSMZ1vMZ2}}
 \end{center}
 \end{figure}
\begin{figure}[!ht]
 \begin{center}
\includegraphics[width=3.15in, height=2.90in]{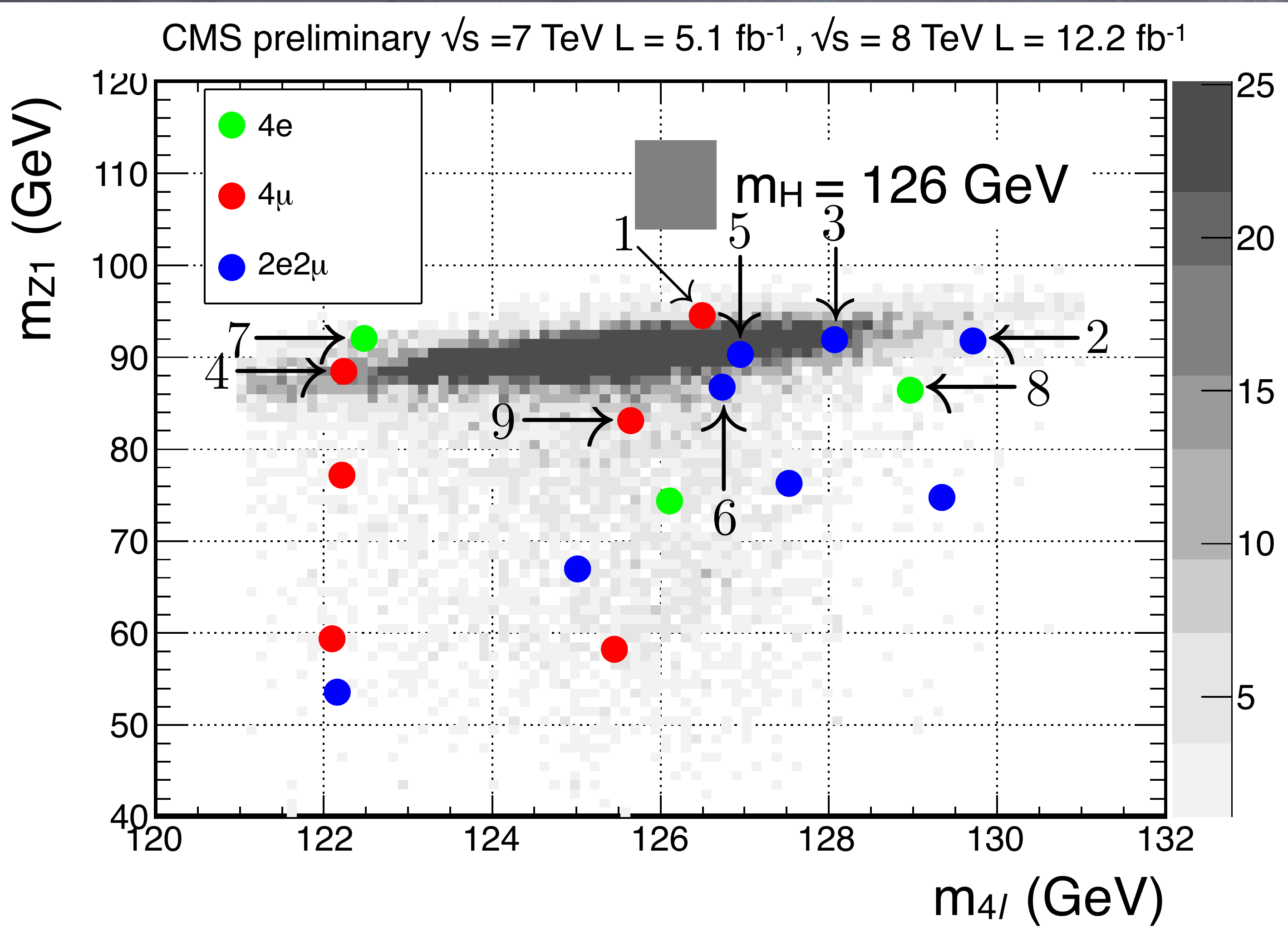}
\includegraphics[width=3.15in, height=2.90in]{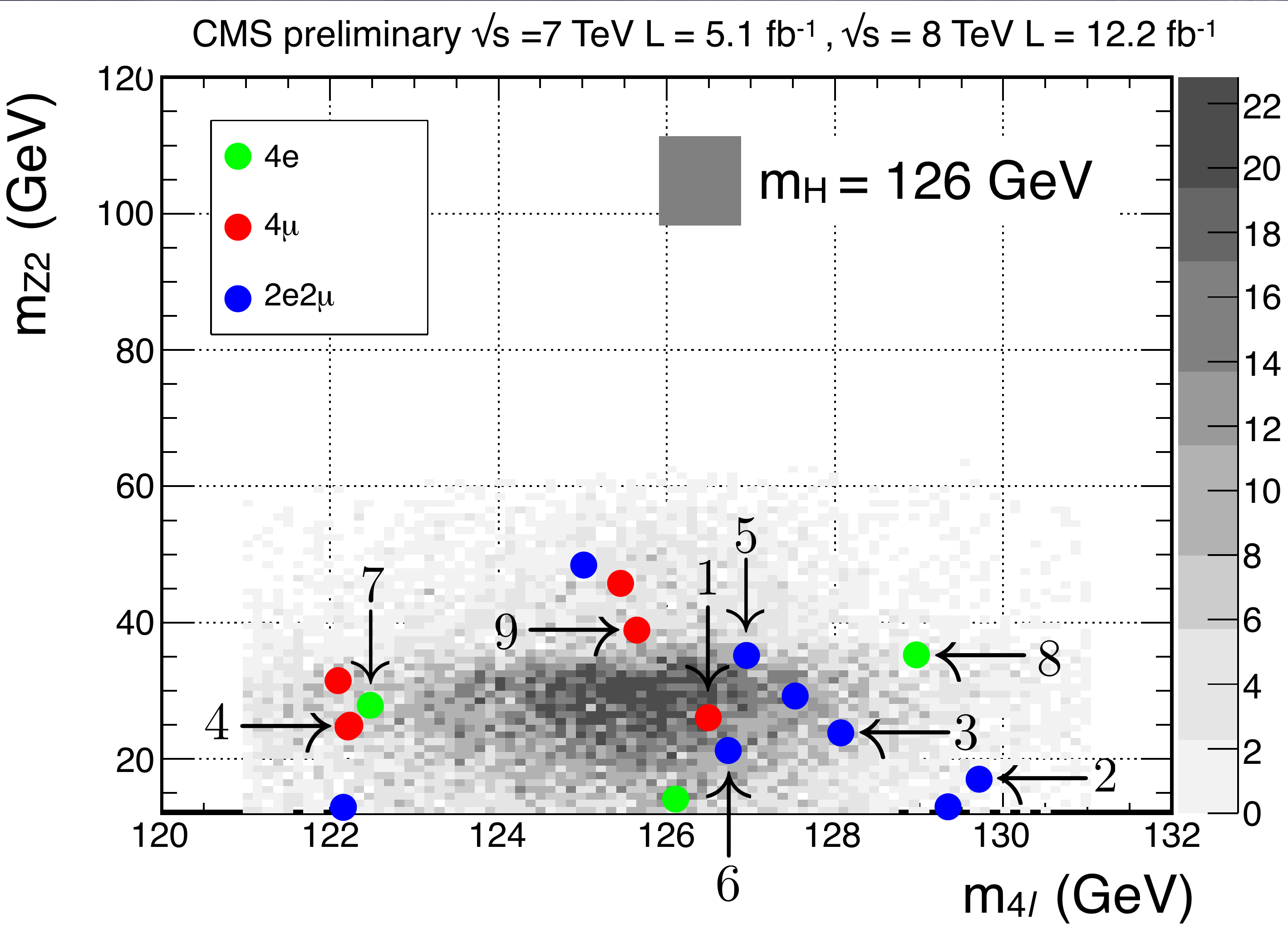}
\vskip 0.2in
\caption{The Dalitz plot of $M_{Z1}$ (left) and $M_{Z2}$ (right)
  vs.~$M_{4\ell}$ in the region $120\,\gev < M_{4\ell} < 130\,\gev$ from
  CMS~\cite{CMSHigtwiki}. The numbering of events is the same as in
  Fig.~\ref{fig:CMSMZ1vMZ2} and is described in the text.
  \label{fig:CMSMZvM4l}}
 \end{center}
 \end{figure}

\begin{enumerate}
  
\item ATLAS and CMS obtained the $\mu(\gamma\gamma) \equiv \sigma B(pp\to X
  \to \gamma\gamma)/\sigma B(pp \to H \to \gamma\gamma) = 1.8\pm 0.7$
  and~$1.6\pm 0.4$ for the SM Higgs $H$. This is the most compelling evidence
  for production of the new particle $X$ and for its interpretation as a
  Higgs boson. This ``signal strength'' and others, $\mu(F)$ for final state
  $F$, are summarized in Fig.~\ref{fig:mufig}. The $\mu(\gamma\gamma)$ is
  dominated by events with no tagged forward jet (untagged), though there is
  some contribution from events with one or more tagged forward jet ---
  so-called vector-boson fusion or VBF tag, though there is no evidence that
  the tagged jet is associated with $WW$ or $ZZ$ fusion of $X$, and it may
  have arisen from gluon ($gg$) fusion. Note that the CMS $\gamma\gamma$ data
  has not been updated since July~2012.
 
\item Despite its low rate, the channel $X \to ZZ^* \to 4\ell$ is very
  important because of its excellent mass resolution. Because of this, it has
  the highest significance after $\gamma\gamma$. Nevertheless, we believe
  that this $ZZ^*$ (and $Z\gamma^*$) data is still subject to rather large
  statistical fluctuations and does not yet provide the evidence for a
  Higgs-boson interpretation of $X$ commonly attributed to it as, e.g., in
  Refs.~\cite{PausHCP,EinsweilerHCP}.

\begin{figure}[!t]
 \begin{center}
\includegraphics[width=3.15in, height=3.15in]{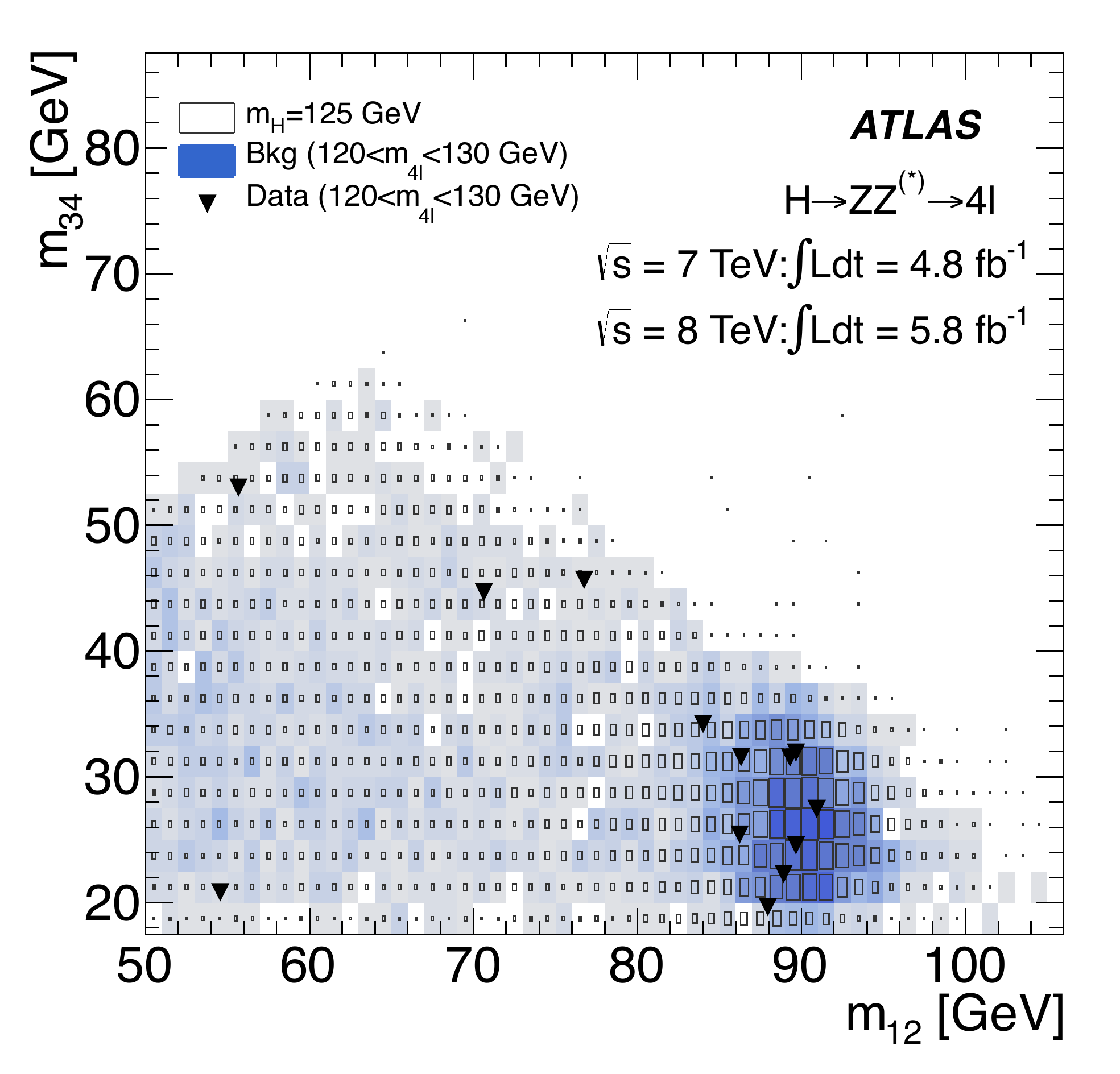}
\includegraphics[width=3.15in, height=3.15in]{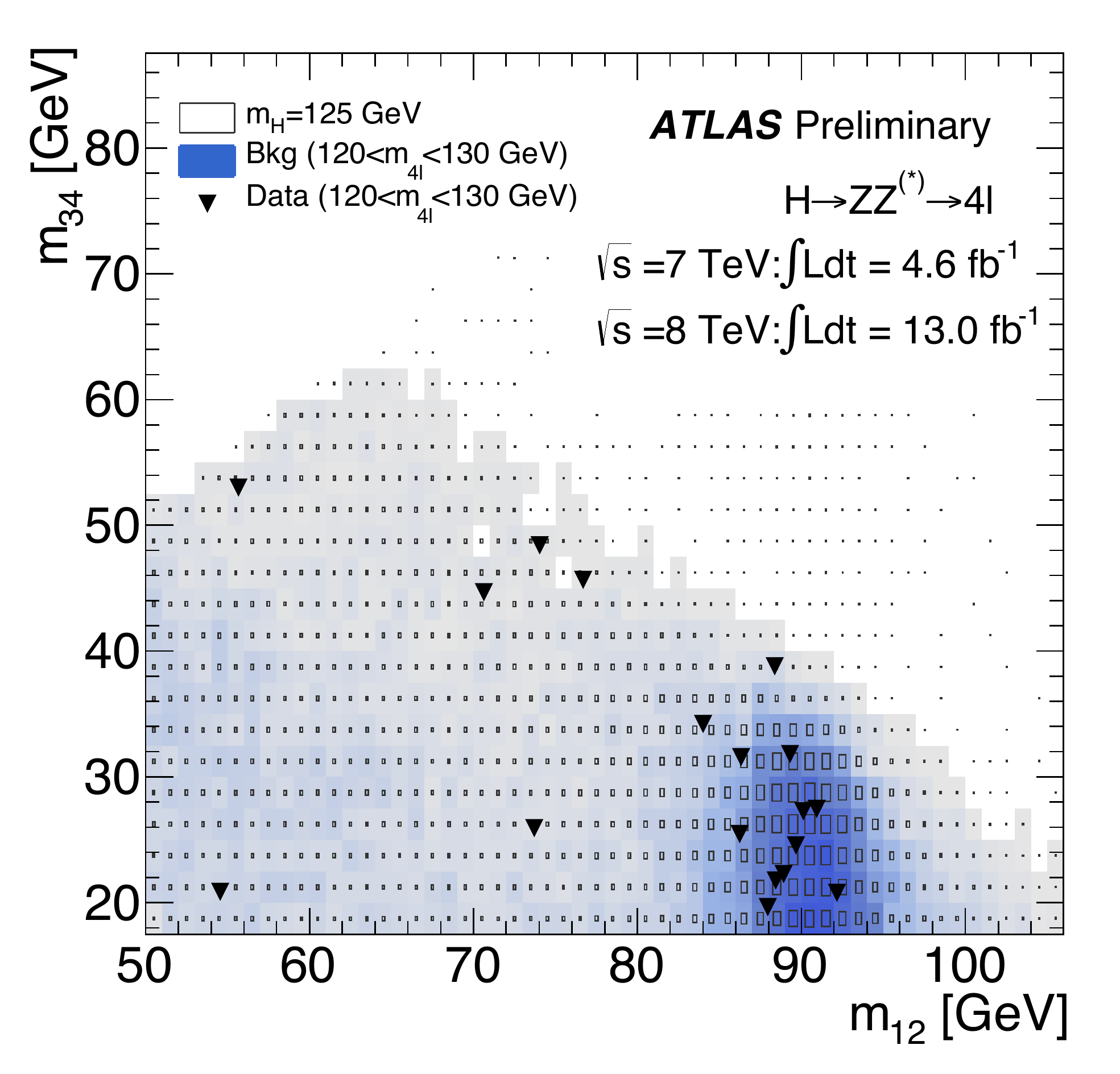}
%\vskip -0.5in
\caption{The Dalitz plot of high vs.~low dilepton mass, $M_{12}$ vs.~$M_{34}$,
  in the region $120\,\gev < M_{4\ell} < 130\,\gev$ from ATLAS in July
  2012~\cite{:2012gk} (left) and December 2012~\cite{ATLASCONFZZ} (right).
  \label{fig:ATLASMZ1vMZ2}}
 \end{center}
 \end{figure}
% 

 %\vfil\eject
 
 In the CMS data released in July, there were ten events, including an
 expected background of three, with four-lepton invariant mass $120\,\gev <
 M_{4\ell} < 130\,\gev$. As seen on the left in Fig.~\ref{fig:CMSMZ1vMZ2},
 only two (or, at most, four) of the events in CMS's plot of $M_{Z1}$
 vs.~$M_{Z2}$ appear to have a real $Z$-boson, those with $85\,\gev \simle
 M_{Z1} \simle 95\,\gev$, whereas~70--80\% of $ZZ^*$ in this mass range are
 expected to have a real~$Z$. This data was based on two sets of about
 $5\,\ifb$ each taken at $\sqrt{s} = 7$ and $8\,\tev$. CMS updated its
 $ZZ^*/\gamma^*$ data in November, with a total of $12.2\,\ifb$ at $8\,\tev$.
 This data has 17~events with an expected background of six in $M_{4\ell} =
 120$--$130\,\gev$. The new $M_{Z1}$ vs.~$M_{Z2}$ plot has 8--9~real $Z$'s,
 i.e., essentially all of the new events are in the dark signal region; see
 Fig.~\ref{fig:CMSMZ1vMZ2}.\footnote{It is unclear to us why the
   $M_{Z1}$-width of this region almost doubled between July and November.
   That did not happen with the ATLAS data in Fig.~\ref{fig:ATLASMZ1vMZ2}.}
 Statistically, CMS was unlucky in July or unlucky in November.
 
 A closer look at the CMS $ZZ^*$ signal data makes it even less convincing.
 In Fig.~\ref{fig:CMSMZ1vMZ2} we numbered the 8~or~9~``golden'' events
 with a real $Z$. Numbers~1 and~2 are the original two golden events. In
 Fig.~\ref{fig:CMSMZvM4l} all the events, including the ones we numbered, are
 shown in two plots, $M_{Z1}$~vs.~$M_{4\ell}$ and $M_{Z2}$~vs.~$M_{4\ell}$,
 from ~\cite{CMSHigtwiki}. In $M_{Z1}$~vs.~$M_{4\ell}$, only events~3 and~5
 are in the Monte Carlo signal's dark region.  Events~1,4,6 are on the
 lighter edges of this region. In $M_{Z2}$~vs.~$M_{4\ell}$, only events~1
 and~6 are the dark part of the signal region; marginally, events~3,5,7 are
 may be included. Thus, {\em no} real-$Z$ event is in the dark signal region
 of all three plots. More generously, only the four real-$Z$ events~1,3,5,6
 are in the signal region of all three plots. This is about 1/2~the expected
 number of $H \to ZZ^* \to 4\ell$ signal events.
 
 The ATLAS $ZZ^*/\gamma^*$ data released in July~\cite{:2012gk} and in
 December~\cite{ATLASCONFZZ} are shown in Fig.~\ref{fig:ATLASMZ1vMZ2}. Note
 first that the ATLAS plots reveal that the region of maximum $H\to ZZ^*$
 production is right where the background peaks, usually a cause for concern.
 ATLAS's July data, based on $4.8\,\ifb$ at $7\,\tev$ and $5.8\,\ifb$ at
 $8\,\tev$ are more Higgs-like than the July CMS data: there are 13~events
 with $120\,\gev < M_{4\ell} < 130\,\gev$, of which 8--9~appear to have a
 real~$Z$ and are in the Higgs signal region of the Monte Carlo. The data
 released in December included $13\,\ifb$ at $8\,\tev$.  They have 18~events,
 but only two new ones are in the Higgs signal region. (It appears that one
 July event's $M_{34}$ decreased from about $32\,\gev$ to $28\,\gev$.) There
 are ten~apparently real-$Z$ events in the signal region on the right in
 Fig.~\ref{fig:ATLASMZ1vMZ2}. We analyzed these as we did the nine CMS
 events. We found that only two are in the signal region of all three plots.
 A more generous definition of the $H\to ZZ^*$ regions yields four in all
 three plots. As for CMS, it appears that statistics are at work here.
 
 ATLAS~\cite{ATLASCONFZZ} and CMS~\cite{:2012br} have also published angular
 distributions or discriminants based on their $ZZ^* \to 4\ell$ events that
 are intended to differentiate between $J^P = 0^+$ and $0^-$ for X(125).
 Given our arguments that neither experiment's $ZZ^*$ data yet has the
 statistical strength required for a demonstration of $H \to ZZ^*$, we do not
 believe that a convincing spin-parity analysis can be made from this data
 set. This view is strengthened by the actual angular distribution data in
 Fig.~18 or Ref.~\cite{ATLASCONFZZ} and Fig.~2 of Ref.~\cite{:2012br}. They
 appear incapable of distinguishing the two cases.

\item The channel $X \to WW^* \to \ell\nu\ell\nu$ channel is also important,
  but not nearly so much as $ZZ^* \to 4\ell$ because of the large missing
  energy and lack of a well-defined discrete mass for its source. The ATLAS
  and CMS data in July and December were mildly inconsistent. The latest
  quoted signal strengths for this channel are $\mu(WW^*) = 1.5\pm 0.6$ for
  ATLAS~\cite{ATLASCONFggZZ} and $0.7 \pm 0.2$ for CMS~\cite{CMStwiki}

\item The decay $H \to \tau^+\tau^-$ is best sought in the associated
  production modes $WH \to \ell\nu\tau\tau$ and $ZH\to \ellp\ellm \tau\tau$
  because of very large background from $Z \to \tau^+\tau^-$. CMS reported
  $\mu(\tau^+\tau^-) = 0.0\pm 0.8$ in July and $0.9\pm 0.5$ in November. This
  result is dominated by $\tau^+\tau^-$ produced with zero or one jet ($gg$
  and/or VB fusion), but with rather large errors; the result for $W/Z X$
  associated production is consistent, but with very large error. ATLAS
  first reported on this mode in November, with $\mu(\tau^+\tau^-) = 0.8 \pm
  0.7$. In short, the evidence for $X \to \tau^+\tau^-$ is weak, but this is
  not surprising given the difficulty of detecting it.
  
\item In July, neither ATLAS nor CMS reported observing the associated
  production mode $WX \to \ell\nu\bar bb$, but this too is not surprising
  given the large backgrounds to this signal at the LHC. The CDF and D\O\ 
  experiments combined their search for $\bar pp \to WH, \,ZH$ with $H \to
  \bar bb$ and claimed a signal consistent with $X(125)$ at the $3.1\,\sigma$
  level~\cite{Aaltonen:2012qt}. This was surprising considering that $S/B <
  1\%$ for the samples used for this channel~\cite{Aaltonen:2012ii}.
  Moreover, as Fig.~\ref{fig:CDFDzero} shows, the broad mass peak is not a
  convincing fit to $M_H = 125\,\gev$ and its significance is greatest at
  $M_{\bar bb} = 135\,\gev$.\footnote{The CDF-D\O\ paper does not make clear
    what correction was made for lost neutrinos and muons in the 40\% of
    $b$-semileptonic decays in $\bar bb$ states. Therefore, the actual $\bar
    bb$ mass peak might be even higher, closer to
    145--150~GeV~\cite{Aaltonen:2011mk}.} In November, CMS reported $\mu(\bar
  bb) = 1.1\pm 0.6$, entirely from $WX \to \ell\nu\bar bb$~\cite{CMSPASHIG}.
  ATLAS still has no signal, with $\mu(\bar bb) = -0.4 \pm
  1.0$.~\cite{ATLASCONFggZZ}.

\end{enumerate}

\noindent Of course, these fluctuations and disagreements may disappear with
more data. For now, they are tantalizing, and alternative interpretations of
$X(125)$ are worth exploring.

\begin{figure}[!t]
 \begin{center}
\includegraphics[width=3.15in, height=3.15in]{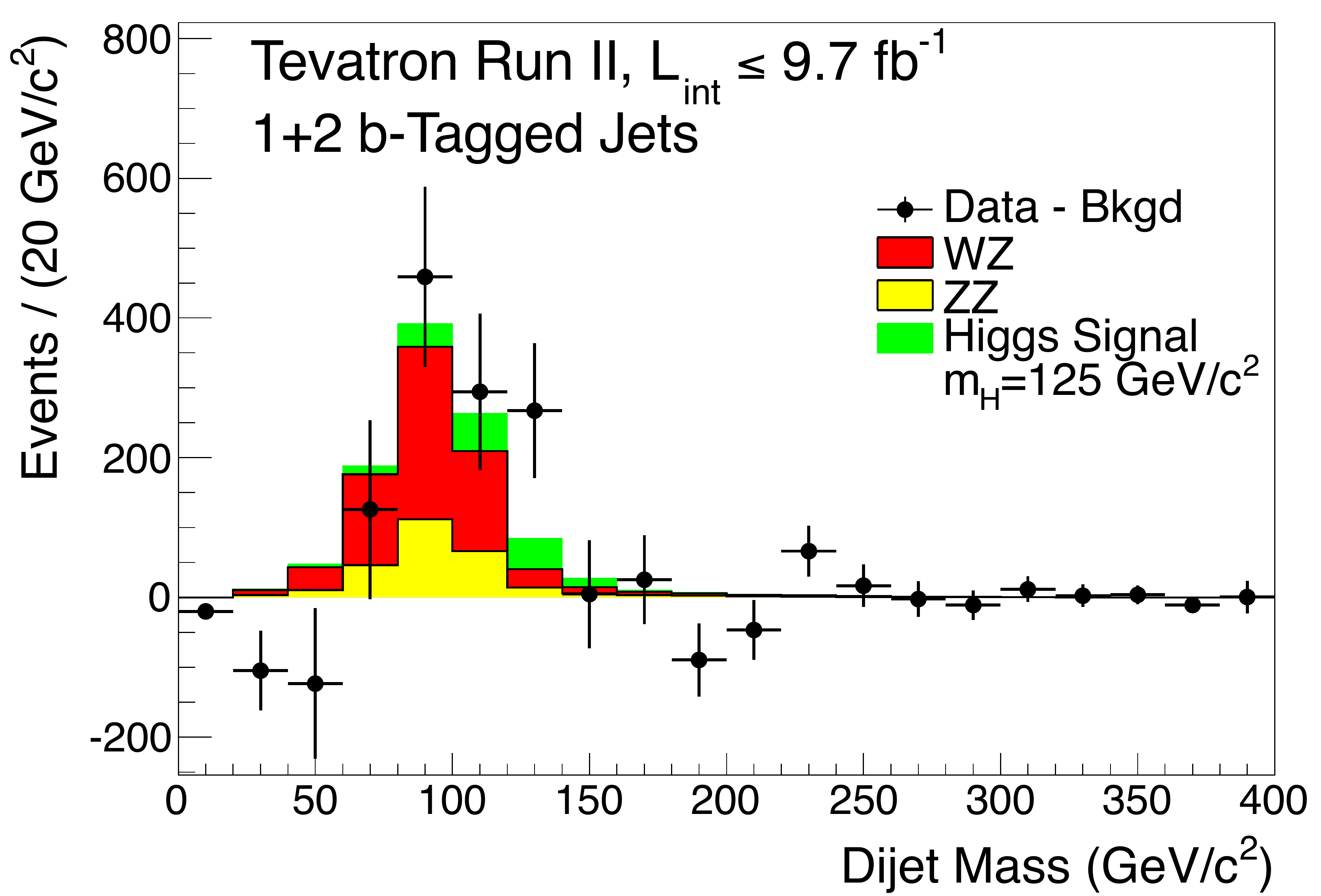}
\includegraphics[width=3.15in, height=3.15in]{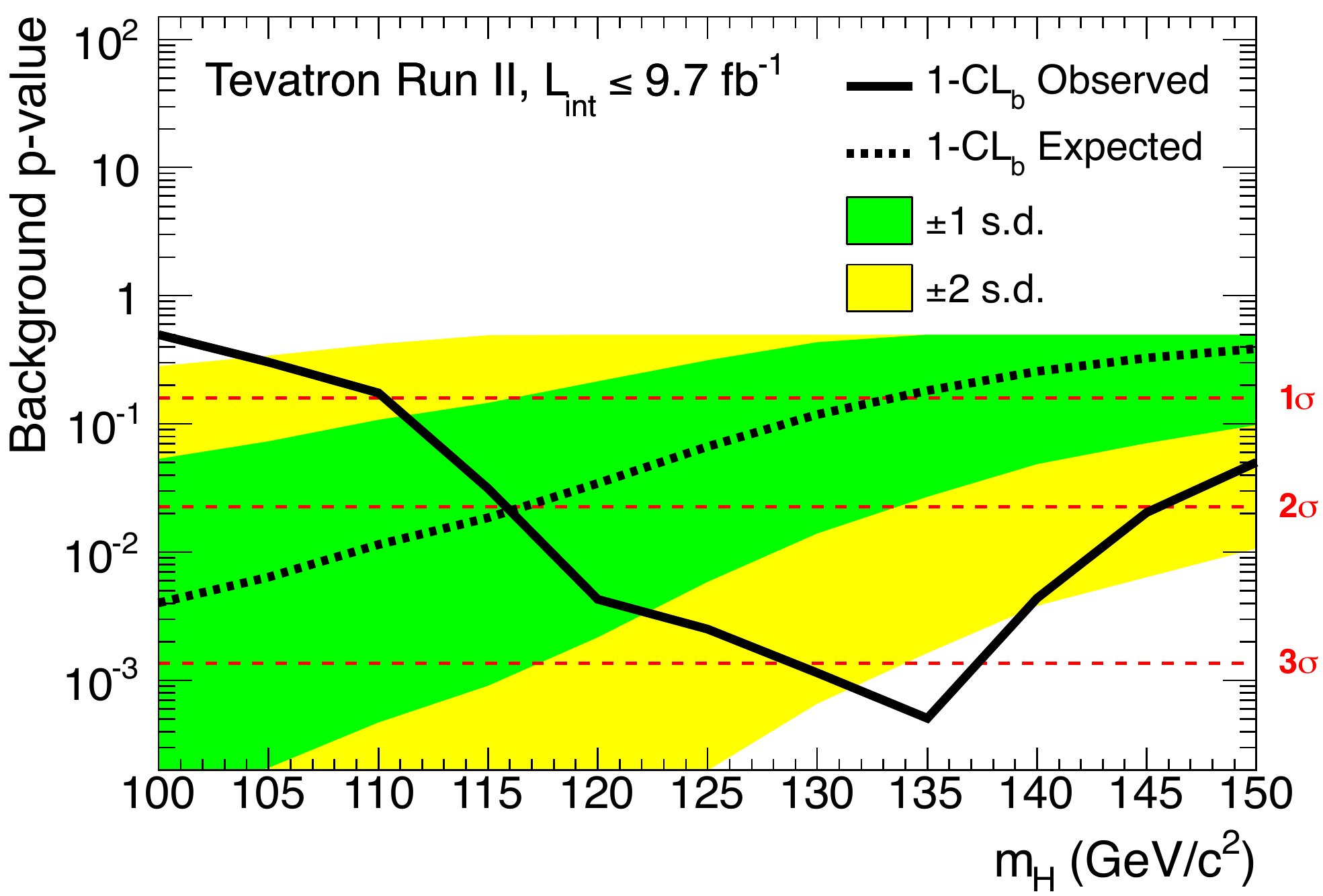}
%\vskip -0.5in
\caption{CDF and D\O\ data on $\bar p p \to W\bar bb$ with $M_{\bar bb}$ in
  the 125~GeV region~\cite{Aaltonen:2012qt}. See the text for comments.
  \label{fig:CDFDzero}}
 \end{center}
 \end{figure}

\section*{3. A Two-Scale Model for the $\etat$}

The technipion $\etat$ is a pseudo-Goldstone boson that must occur in LSTC
models~\cite{Lane:1989ej}. It was referred to as $\tpipr$ in previous papers,
e.g., Ref.~\cite{Lane:2009ct}, which also contains a more complete
description of LSTC. Since $\etat$ is a pseudoscalar and decays to two
photons, it has $CP = -1$. Therefore, it has no renormalizable couplings to a
pair of SM gauge bosons. Its main production mechanism, therefore, must be
via gluon ($gg$) fusion. This requires that $\etat$ be composed, at least in
part, of technifermions carrying ordinary $SU(3)_C$ color.\footnote{The top
  quark cannot couple strongly to $\etat$ nor any other $\tpi$ because, in
  ETC models with fermion-bilinear anomalous dimension $\gamma_m \le
  1$~\cite{Cohen:1988sq}, $m_t$ must arise from some other strong
  interaction, such as topcolor~\cite{Hill:1994hp}.} In LSTC, we usually
assume that the lightest (lowest-scale) technifermions are
$SU(3)_C$-singlets, and we make that assumption here. Thus, for an $\etat gg$
interaction to occur, the higher-scale technifermions must be
colored.\footnote{An alternative in which the lightest-scale technifermions
  are colored might be interesting, but we shall not consider it here. As
  Eq.~(\ref{eq:sinchi}) indicates, this tends to imply a larger value of the
  LSTC parameter $\sin\chi$ in Eq.~(\ref{eq:sinchi}) and that is disfavored
  experimentally~\cite{Collaboration:2012kk,Eichten:2012hs}.} To describe
this, we adopt the following two-scale model:
\bea\label{eq:twoscale}
%\begin{align}
\text{Scale 1:}\quad T_1 \equiv \left(\begin{array}{c}U_1\\
    D_1\end{array}\right)
& = \left\{ \begin{array}{cc}T_{1L} &  = (\fund ,
    1, 2)_{Y_1} \\ 
U_{1R} & = (\fund , 1, 1)_{Q_{U1}} \\
D_{1R} & = (\fund, 1, 1)_{Q_{D1}} 
\end{array}\right. \nn \\
\text{Scale 2:}\quad T_2 \equiv \left(\begin{array}{c}U_2\\
    D_2\end{array}\right)
& = \left\{ \begin{array}{cc} T_{2L} &=
    (\antisymm, \fund , 2)_{Y_2} \\
U_{2R} & = (\antisymm, \fund , 1)_{Q_{U2}} \\ 
D_{2R} & = (\antisymm, \fund , 1)_{Q_{D2}}
\end{array}\right.
%\end{align}
\eea
%
%% % %
%% \bea\label{eq:twoscale}
%% %\begin{align}
%% \text{Scale 1:}\quad T_1 \equiv \left(\begin{array}{c}U_1\\
%%     D_1\end{array}\right)
%% &=& \left\{ \begin{array}{cc}T_{1L} &  = ({\bf N_{TC}} ,
%%     1, 2)_{Y_1} \\ 
%% U_{1R} & = ({\bf N_{TC}} , 1, 1)_{Q_{U1}} \\
%% D_{1R} & = ({\bf N_{TC}}, 1, 1)_{Q_{D1}} 
%% \end{array}\right. \nn \\
%% \text{Scale 2:}\quad T_2 \equiv \left(\begin{array}{c}U_2\\
%%     D_2\end{array}\right)
%% &=& \left\{ \begin{array}{cc} T_{2L} &=
%%     ({\bf \thalf N_{TC}(N_{TC}-1)}, {\bf 3_C} , 2)_{Y_2} \\
%% U_{2R} & = ({\bf \thalf N_{TC}(N_{TC}-1)}, {\bf 3_C} , 1)_{Q_{U2}} \\ 
%% D_{2R} & = ({\bf \thalf N_{TC}(N_{TC}-1)}, {\bf 3_C} , 1)_{Q_{D2}}
%% \end{array}\right.
%% %\end{align}
%% \eea
%% % %
under $(SU(\Ntc),SU(3)_C,SU(2))_{U(1)}$. Here, $Y_i = \thalf(Q_{Ui} + Q_{Di})$.

We emphasize that this model's purpose is to illustrate our LSTC proposal to
account for the $X(125)$ data. Different TC representations and/or input
parameters $(\Ntc$, etc.) could give quantitatively different results, and
more data may require a refinement of the model. Nevertheless, we believe the
model's general features---the interactions $\etat$ has with ordinary matter
and the typical strength of these interactions---will survive as long as the
viability of an LSTC impostor of $X(125)$ does.

When the technifermions $T_1$ and $T_2$ condense, there are a number of
Goldstone bosons (all but three of which must get mass from ETC
interactions~\cite{Eichten:1979ah}) including two color-singlets with $I^G
J^{PC} = 0^+ 0^{-+}$ we call $ \eta_1$ and $\eta_2$. These couple to the
$U(1)$ axial vector currents $j_{i,5\mu} = \half \bar T_i \gamma_\mu \gamma_5
T_i$ as
\be\label{Uone}
\langle\Omega|j_{1,5\mu}|\eta_1(p)\rangle = iF_1 p_\mu, \qquad
\langle\Omega|j_{2,5\mu}|\eta_2(p)\rangle = i\sqrt{3}F_2 p_\mu,
\ee
where $F_1$ and $F_2$ are the basic (canonically normalized) $\tpi$ decay
constants of scales~1 and~2. They are related to the weak decay constant
$F_\pi \equiv v = 246\,\gev$ and the LSTC mixing angle parameter
$\sin\chi$~\cite{Lane:2002sm,Lane:2009ct} by
\be\label{eq:sinchi}
F_\pi = \sqrt{F_1^2 + 3 F_2^2}, \qquad
F_1 = F_\pi\sin\chi, \quad \sqrt{3} F_2 = F_\pi\cos\chi\,.
\ee
A recent search by CMS for $\tro \to WZ \to 3\ell\nu$ put a 95\% upper limit
of about $20\,\fb$ on its cross section at $M_{\tro} = 275$--$290\,\gev$ and
$M_{\tpi} > 140\,\gev$~\cite{Collaboration:2012kk}.  This requires $\sin\chi
\simle 0.30$ for the LSTC model with these masses~\cite{Eichten:2012hs}.
While this bound is relevant for the case of little or no $\etat$-$\tpiz$
mixing, sizable mixing probably weakens it; see Sec.~7.

The $U(1)$ currents have divergences with TC-gluon anomalous terms and other
explicit breaking:
\be\label{eq:divj}
\partial^\mu j_{i,5\mu} = -\frac{g^2_{TC}}{16\pi^2} N_i G_{T,\mu\nu}
\widetilde G_T^{\mu\nu} + i[\CQ_{i,5},\CH_{ETC}] + \cdots\,,
\ee
where $\widetilde G_{T,\mu\nu} = \thalf \epsilon_{\mu\nu\lambda\rho}
G_T^{\lambda\rho}$, $\CQ_{i,5} = \int d^3x\, j_{i,50}$, $\CH_{ETC}$ is a
4-technifermion interaction involving $T_1$ and $T_2$, and the ellipses are
$SU(3)_C \otimes SU(2) \otimes U(1)$ anomalous divergences that will be
specified in Sec.~5.  In Eq.(\ref{eq:divj}) the numerical factors are $N_i =
2T(R_{TC,i}) d(R_{C,i})$, where the factor~2 is for isodoublet
technifermions, $T(R_{TC})$ is the trace of a square of generators for
TC-representation $R$ ($=\half$ for fundamentals of $SU(\Ntc)$) and $d(R_C)$
is the dimension of the $SU(3)_C$ representation. In the model of
Eq.~(\ref{eq:twoscale}),
\be
N_1 = 2\cdot\thalf\cdot 1 = 1,\quad
N_2 = 2\cdot\thalf(\Ntc-2)\cdot 3 = 3(\Ntc -2)\,.
\ee
The current $j'_{5\mu} = j_{1,5\mu} + j_{2,5\mu}$ is conserved by ETC
interactions (see Sec.~4) but not by the TC anomaly:
\be\label{divjprime}
\partial^\mu j'_{5\mu} = -\frac{g^2_{TC}}{16\pi^2} (N_1 + N_2) G_{T,\mu\nu}
\widetilde G_T^{\mu\nu} +\cdots\,.
\ee
It couples to a linear combination $\eta_T'$ of $\eta_1$ and $\eta_2$ which
gets its mass mainly from TC instantons and is heavy. The orthogonal linear
combination is the $\etat$ and its mass arises from $\CH_{ETC}$. It couples
to the TC-anomaly-free current
\bea\label{eq:jetaT}
j_{5\mu} &=& N_2 j_{1,5\mu} - N_1 j_{2,5\mu} \\
\partial^\mu j_{5\mu} &=& i[N_2 \CQ_{1,5} - N_1 \CQ_{2,5}, \CH_{ETC}] + \cdots  
= i(N_1 + N_2) [\CQ_{1,5},\CH_{ETC}] +\cdots\,.
\eea

Let us write
\bea\label{eq:etamix}
|\eta_T'\rangle &=&|\eta_1\rangle \sin\eta + |\eta_2\rangle \cos\eta \nn\\
|\etat\rangle &=&|\eta_1\rangle \cos\eta - |\eta_2\rangle \sin\eta \,.
\eea
The mixing angle $\eta$ is determined by noting that, unless the matrix
element $\langle\Omega|j_{5\mu}|\eta_T'\rangle = 0$ in the limit $\CH_{ETC}
\to 0$, then $M_{\eta_T'} \cong 0$ since this current is TC-anomaly
free. This yields
\be\label{eq:sineta}
 \sin\eta = \frac{\sqrt{3} N_1 F_2}{\Fetat},\quad
   \cos\eta = \frac{N_2 F_1}{\Fetat},\qquad
{\rm where}\quad
 \Fetat = \sqrt{N_2^2 F_1^2 + 3 N_1^2 F_2^2}\,.
\ee
Noting that
\be\label{eq:FetaFpi}
\Fetat = \sqrt{N_2^2 \sin^2\chi + N_1^2 \cos^2\chi}\, F_\pi\,,
\ee
we have
\be\label{eq:sinetachi}
  \sin\eta = \frac{N_1\cos\chi}{\sqrt{N_2^2 \sin^2\chi + N_1^2
     \cos^2\chi}},\quad
  \cos\eta = \frac{N_2\sin\chi}{\sqrt{N_2^2 \sin^2\chi + N_1^2
      \cos^2\chi}}\,.
\ee
For $\Ntc = 4$ and $\sin\chi = 0.3$, we have $N_1 = 1$, $N_2 = 6$, $\sin\eta
= 0.468$, $\cos\eta = 0.884$, and $\Fetat = 501\,\gev$ is the normalized decay
constant of the $\etat$. 

\section*{4. $\etat$-$\tpiz$ Mixing}

The state $\etat$ discussed in Sec.~3 generally is not a mass eigenstate. In
the model we have presented and in similar ones, the ETC interactions that
give it mass also mix it with the neutral isovector technipion $\tpiz$
discussed in Refs.~\cite{Eichten:2011sh, Eichten:2012hs}.  This effects not
only $\etat$ phenomenology but, as we discuss in Sec.~7, the LSTC description
of the CDF dijet excess observed near $\Mjj = 150\,\gev$ in $Wjj$
production~\cite{Aaltonen:2011mk,CDFnew}. The ETC interactions of $T_1$ and
$T_2$ must be $SU(\Ntc)\otimes SU(3)_C \otimes SU(2) \otimes U(1)$ invariant.
For our model they have the following form at energies far below the masses
$M_{1,2,3}$ of ETC gauge bosons:
\bea\label{eq:TTETC}
\CH_{ETC}  &=& \frac{\getc^2}{M_1^2}\, \bar T_{1L}\gamma^\mu T_{1L}\bar
T_{1R}\gamma_\mu (a_1 + b_1\tau_3) T_{1R}\nn\\
&+&  \frac{\getc^2}{M_2^2}\, \left(\bar
  T_{1L}\gamma^\mu T_{2L}\bar T_{2R}\gamma_\mu (a_2 + b_2\tau_3) T_{1R} +
  {\rm h.c.}\right) \nn \\
&+& \frac{\getc^2}{M_3^2}\, \bar T_{2L}\gamma^\mu T_{2L}\bar T_{2R}\gamma_\mu
(a_3 + b_2\tau_3) T_{2R}\,.
\eea
The $SU(\Ntc) \otimes SU(3)_C$ indices of these interactions are suppressed,
but the structure of the middle term, e.g., is
\be\label{eq:indices}
\bar T^{\alpha}_{1L}\gamma^\mu T^{[\alpha\beta],k}_{2L}\bar
T^{[\beta\gamma],k}_{2R}\gamma_\mu (a_2 + b_2\tau_3) T^{\gamma}_{1R}\,,
\ee
where $\alpha,\beta,\gamma = 1,2,\dots,\Ntc$ are $SU(\Ntc)$ indices with
$[\alpha\beta] = -[\beta\alpha]$ and $k= 1,2,3$ is an $SU(3)_C$ index. The
$SU(2)_R$ violation in the $b$-terms is necessary to split up from
down-fermions. We expect $a_i, |b_i| = \CO(1)$ with $a_i > 0$ while $b_i$
may have either sign.

To a very good approximation, the masses and mixing of the technipions
$\tpipm$, $\tpiz$ and $\etat$ come entirely from the $\bar T_1 T_2 \bar T_2
T_1$ terms, and they are determined as follows: In the absence of
$\etat$-$\tpiz$ mixing, the mass eigenstates $|\pi_T^a\rangle$ ($a=1,2,3)$
are the linear combination
\be\label{eq:tpistate}
|\pi_T^a\rangle = \cos\chi|\pi_1^a\rangle - \sin\chi|\pi_2^a\rangle,\,
\ee
where $|\pi_{1,2}^a\rangle$ are the scale-1,2 color-singlet technipions. The
mixing angle $\chi$ was defined in Eq.~(\ref{eq:sinchi}), with $\sin\chi >
0$. The orthogonal combinations are the three Goldstone components of the
electroweak bosons, $|W_L^a\rangle$. The state $|\pi_T^a\rangle$ does not
couple to the conserved electroweak axial current $j^{a,EW}_{5\mu} =
j^a_{1,5\mu} + j^a_{2,5\mu} + \cdots$, where $j^a_{i,5\mu} = \thalf \bar
T_i\gamma_\mu \gamma_5 \tau_a T_i$; if it did, $M_{\tpi} = 0$. The $\pi_T^a$
current we will use for calculating $M_{\tpi}$ is
\be\label{eq:picurr}
j^a_{5\mu} = j^a_{1,5\mu} \cot\chi - j^a_{2,5\mu} \tan\chi\,.
\ee
This current couples to $\tpi$ in Eq.~(\ref{eq:tpistate}) with strength $F_\pi$,
\be\label{eq:interp} \langle\Omega|j^a_{5\mu}|\pi^b_T(p)\rangle =
 iF_{\pi} p_\mu \delta_{ab}\,,  \ee
 but not to the orthogonal combination, the erstwhile Goldstone bosons that
 are the longitudinally-polarized $W^\pm$ and $Z$. Then, with $\CQ_5^a =
 \int d^3x\, j^a_{50}$ for $a=1,2,3$, and using isospin and parity invariance
 of the vacuum state $\rvac$, we obtain~\cite{Dashen:1969eg}
\bea\label{eq:tpimass} F_{\pi}^2 M^2_{\tpi} &=&
i^2 \lvac[\CQ_5^a,[\CQ_5^a,\CH_{ETC}]]\rvac \nn\\
&=& \frac{i^2 a_2 \getc^2}{2 M_2^2 \sin^2\chi\cos^2\chi} \lvac \bigl[\bar
T_{1L} \gamma^\mu \tau_a T_{2L} \bar T_{2R} \gamma_\mu \tau_a T_{1R} %\nn\\
+ \bar T_{1L} \gamma^\mu T_{2L} \bar T_{2R} \gamma_\mu T_{1R} + {\rm h.c.}
\bigr]\rvac \nn\\
&=& \frac{2 i^2 a_2\getc^2}{M_2^2\sin^2\chi\cos^2\chi} \lvac \bigl[\bar
T_{1L} \gamma^\mu T_{2L} \bar T_{2R} \gamma_\mu T_{1R}\bigr]\rvac \,, \eea
Similarly, with $\CQ_5 = N_2 \CQ_{1,5} - N_1 \CQ_{2,5}$, we get
\bea\label{eq:etapimass}
\Fetat^2 M^2_{\etat} &=&  \left[(N_1 + N_2) \sin\chi\cos\chi \, F_\pi
M_{\tpi}\right]^2  \\
F_\pi\Fetat M^2_{\etat \tpiz} &=& (b_2/a_2)(N_1 + N_2)
 \sin\chi\cos\chi\, F_\pi^2 M^2_{\tpi}\,. \\
\eea
Then, using Eq.~(\ref{eq:FetaFpi}) for $\Fetat$,
\bea\label{eq:massratios}
\left(\frac{M_{\etat}}{M_{\tpi}}\right)^2 &=& \frac{((N_1 + N_2)
\sin\chi\cos\chi)^2}{N_1^2 + (N_2^2 - N_1^2)\sin^2\chi} = 0.967\,(0.998)\,,\\
\left(\frac{M_{\etat\tpiz}}{M_{\tpi}}\right)^2 &=&
    \frac{b_2}{a_2} \left(\frac{M_{\etat}}{M_{\tpi}}\right)^2\,.
\eea
Here, $M_{\tpi}$ is the mass of the charged $\tpipm$, which is unaffected by
the $|\Delta I| = 1$ isospin breaking in $\CH_{ETC}$. The numerical values in
Eq.~(\ref{eq:massratios}) are for $\sin\chi = 0.30$ and $\Ntc = 4\, (6)$.
They will be close to one when $(N_2\sin\chi)^2 \gg N_1^2$ and $\sin^2\chi
\ll 1$, as it is here.

Thus, in two-scale models like the one presented here, we have the surprising
result that the mass eigenstates are nearly 50-50 admixtures of the neutral
isoscalar and isovector technipions,
\bea\label{eq:etapimix}
|\etal\rangle &\cong& \sqrt{\thalf}\left(|\eta_T\rangle -
  {\rm sgn}(b_2)\,\tpiz\rangle\right),
\nn\\ 
|\etah\rangle &\cong& \sqrt{\thalf}\left(|\eta_T\rangle +
  {\rm sgn}(b_2)\,\tpiz\rangle\right),
\eea
with masses
\be\label{eq:etamasses}
M_{\etal} \cong M_{\tpi}\sqrt{1 - |b_2|/a_2},\,\quad 
M_{\etah} \cong M_{\tpi}\sqrt{1 + |b_2|/a_2}.
\ee

How do we determine the mass of $\etah$? One way is this: In recent
work~\cite{Eichten:2011sh,Eichten:2012hs} we ascribed the CDF dijet mass
excess near $150\,\gev$~\cite{Aaltonen:2011mk,CDFnew} to the production and
decay of the lightest isovector technipions, produced in the LSTC process
$\tro \to W\tpi \to \ell\nu jj$. In the present framework, we assume that
what CDF saw was $\troz \to W^\pm \tpimp$, with $M_{\tpipm} =
150$--$160\,\gev$. The $\tpiz$ is now part of the mixed-state $\etal$, our
Higgs impostor, observed by ATLAS and CMS with mass $125\,\gev$. Then, from
Eq.~(\ref{eq:etamasses}), $M_{\etah} = 170$--$190\,\gev$. In Sec.~7, we will
see how this interpretation alters LSTC phenomenology at the LHC.\footnote{We
  estimate that the Tevatron rate for $W^\pm \tpimp$ production is about
  $2.4\,\pb$, essentially the same as our prediction of the total $W\tpi$
  rate in Ref.~\cite{Eichten:2011sh}. This estimate is rough because the {\sc
    Pythia} code~\cite{Sjostrand:2006za} does not properly describe the model
  with $\etat$-$\tpiz$ mixing; see Sec.~7.}  This rather precise prediction
for $M_{\etah}$ is satisfying, but it does rely on our description of the CDF
excess. If we gave up that description, we would still expect that the
$\etah$---a pseudo-Goldstone boson composed mainly of lighter scale
technifermions---would not be very much heavier than $\etal$. This is clear
from Eqs.~(\ref{eq:etamasses}) so long as $|b_2|/a_2$ is not close to
one. The converse expectation is also likely true: If $X(125)$ is to be
interpreted as an $\etat$ of low-scale technicolor, then there are other
technihadron states nearby, and they should be accessible in hadron collider
experiments.

\section*{5. $\etat$ and $\tpiz$ Interactions}

The couplings between the $CP$-odd $\etat$ and a pair of SM gauge bosons or
SM fermion-antifermion pairs ($\bar ff$) are given by
\bea\label{eq:SMcouplings}
\CL_{\etat} &=& \frac{\etat}{\Fetat} \partial^\mu j_{5\mu} \equiv 
\frac{\etat}{\Fetat} \partial^\mu \left(N_2 j_{1,5\mu} - N_1
  j_{2,5\mu}\right)\nn  \\
&=& {\rm SM\,\,gauge\,\, boson\,\, anomaly\,\, terms} + i[\CQ_5, \CH_{ETC}]\,.
\eea
A similar expression holds for $\tpiz$ with $F_{\tpi} \equiv F_\pi$.

%% The anomaly terms are contained in the gauged WZW interaction calculated in
%% Refs~\cite{Wess:1971yu,Witten:1983tw}. For interactions involving chiral
%% gauge groups, the simplest way to calculate them is to expand the WZW term to
%% linear order in the technipion fields.  We can use the formalism of
%% Refs.~\cite{Wess:1971yu,Witten:1983tw} by introducing a nonlinear-sigma
%% formulation of our model, specifically,
The anomaly terms are obtained as was the gauged WZW interaction in
Refs~\cite{Wess:1971yu,Witten:1983tw}. For chiral
gauge groups, the simplest way to calculate them is to expand the WZW term to
linear order in the technipion fields using a nonlinear-sigma
formulation of our model,
\be\label{eq:nonlinear}
\Sigma_1 = \exp\Big(\frac{2i{\bs \pi}_1}{F_1}\Big)\,, \qquad
\Sigma_2 = \exp\Big(\frac{2i{\bs \pi}_2}{\sqrt{3} F_2}\Big)\,,
\ee
with covariant derivative $D_{\mu} \Sigma_{i} = \partial_{\mu} \Sigma_i - i
\mathcal A_L \Sigma_i + i\Sigma_i \mathcal A_R$ where $\mathcal A_L =
\thalf(g W_\mu^a \tau_a  + g' Y_i\, B_{\mu} \tau_0)$ and
$ \mathcal A_R = \thalf g' B_{\mu}(\tau_3 + Y_i\tau_0)$; ${\bs
  \pi}_i = \thalf(\pi_i^a \tau_a + \eta_i\tau_0)$, with $\tau_0 = {\bs 1}_2$
%% %
%% \bea\label{eq:covariant}
%% D_{\mu} \Sigma_{i} &=& \partial_{\mu} \Sigma_i - i \mathcal A_L \Sigma_i + i
%% \Sigma_i \mathcal A_R\,, \nn \\ 
%% \mathcal A_L &=& \thalf\left(g W_\mu^a \tau_a  + g' Y_i\, B_{\mu}
%% \tau_0\right)\,, \nn \\ 
%% \mathcal A_R &=& \thalf g' B_{\mu}\left(\tau_3 + Y_i\tau_0\right)\,.
%% \eea
%% %
%% Here, ${\bs \pi}_i = \thalf(\pi_i^a \tau_a + \eta_i\tau_0)$, where $\tau_0
%%
and $F_1, F_2$ are the scale--1,2 technipion decay constants defined 
earlier. Applying this setup to Eq.~(69) of Ref.~\cite{Harvey:2007ca}, each
techni-sector contributes a WZW term weighted by a coefficient that depends
on the number of degrees of freedom in that sector. The total WZW interaction
is then $\mathcal L_{WZW} = \mathcal L_{WZW,1} + \mathcal L_{WZW,2}$.  For
$\eta_T, \pi_T$ interactions involving vectorial gauge groups, such as
$\eta_T \ra \gamma \gamma$ or $\eta_T \ra g g$, the WZW result has the
familiar form,
\be\label{eq:anom}
\partial^\mu j_{i,5\mu} = -\frac{g^2_A}{32\pi^2} 
{\rm Tr}\left(\tau_0\, \{t^A_{i,a}, t^A_{i,b}\}\right)\,
G^{Aa}_{\mu\nu} \widetilde G^{Ab,\mu\nu}\,,
\ee
where $t^A_{i,a}$ is the $a$-th generator of technifermion doublet $T_i$ in
gauge group $A$. The corresponding expression for $\partial^\mu j^3_{5\mu}$
has the trace ${\rm Tr}(\tau_3\, \{t^A_{i,a}, t^A_{i,b}\})$.

Since only the isoscalar $\eta_2$ couples strongly to $SU(3)_C$ gluons
through a loop of the color-triplet $T_2$-fermions (see footnote~3), we have
\be\label{eq:etagg}
\CL_{\etat gg} = \sqrt{2}\CL_{\eta_{L,H} gg} =
\frac{g^2_C}{64\pi^2\Fetat} \left[N_1 \Ntc(\Ntc-1)\right] \etat
G^\alpha_{C,\mu\nu} \widetilde G_C^{\alpha,\mu\nu}\,.
\ee
Because of the large numerator, $\CL_{\etat gg}$ is stronger that the
standard $H$ coupling to two gluons.

The nonzero WZW couplings of $\etat$ and $\tpiz$ to a pair of $SU(2)\otimes
U(1)$ bosons are
\bea\label{eq:etapiWZW}
\CL_{\etat BB} &=& -\frac{g'^2\Ntc}{96\pi^2\Fetat}\biggl[N_2(1 + 12 Y_1^2) 
 -\tthalf N_1 (\Ntc-1)(1 + 12 Y_2^2)\biggr]\etat B_{\mu\nu} \widetilde
B^{\mu\nu}\,,\\
\CL_{\etat WW} &=& -\frac{g^2\Ntc}{96\pi^2\Fetat}\biggl[N_2 -\tthalf N_1
  (\Ntc-1)\biggr]\etat W^a_{\mu\nu} \widetilde W^{a,\mu\nu}\,,\\
\CL_{\etat WB} &=& -\frac{gg'\Ntc}{96\pi^2\Fetat}\biggl[N_2 -\tthalf N_1
  (\Ntc-1)\biggr]\etat W^3_{\mu\nu} \widetilde B^{\mu\nu}\,,\\
\CL_{\tpiz BB} &=& -\frac{g'^2\Ntc}{16\pi^2 F_\pi} \biggl[Y_1 \cot\chi
 -\tthalf (\Ntc-1)Y_2 \tan\chi \biggr]\tpiz B_{\mu\nu} \widetilde B^{\mu\nu}\,,\\
\CL_{\tpiz WB} &=& -\frac{gg'\Ntc}{16\pi^2 F_\pi}\biggl[Y_1 \cot\chi
-\tthalf (\Ntc-1)Y_2 \tan\chi\biggr]\tpiz W^3_{\mu\nu} \widetilde B^{\mu\nu}\,.
\eea
From these we obtain
\bea\label{eq:etaggZZWW}
\CL_{\etat\gamma\gamma} &=& -\frac{e^2\Ntc}{32\pi^2\Fetat}
\biggl[N_2(1+ 4 Y_1^2) -\tthalf N_1(\Ntc-1)(1+ 4 Y_2^2)\biggr]
\etat F_{\mu\nu} \widetilde F^{\mu\nu}\,,\\
\CL_{\etat Z\gamma} &=& -\frac{e\sqrt{g^2+g'^2}\Ntc}{32\pi^2\Fetat}
\biggl[N_2(1 - 2(1+4Y_1^2)\sin^2\thw) \nn\\
 && \quad -\tthalf N_1 (\Ntc-1)(1 - 2(1+4Y_2^2)\sin^2\thw) \biggr]
\etat F_{\mu\nu} \widetilde Z^{\mu\nu}\,,\\
\CL_{\etat ZZ} &=& -\frac{(g^2+g'^2)\Ntc}{96\pi^2\Fetat}
\biggl\{N_2\biggl[1 - 3\sin^2\thw + 3(1+4Y_1^2)\sin^4\thw\biggr] \nn\\
&&  \quad -\tthalf N_1 (\Ntc-1)\biggl[1 - 3\sin^2\thw +
  3(1+4Y_2^2)\sin^4\thw\biggr]\biggr\}
\etat Z_{\mu\nu} \widetilde Z^{\mu\nu}\,,\\
\CL_{\etat W^+W^-} &=& -\frac{g^2\Ntc}{48\pi^2\Fetat}\biggl[N_2 -\tthalf N_1
  (\Ntc-1)\biggr] \etat W^+_{\mu\nu} \widetilde W^{-,\mu\nu}\,,\\
\CL_{\tpiz\gamma\gamma} &=&  -\frac{e^2\Ntc}{8\pi^2 F_\pi}
\biggl[Y_1\cot\chi - \tthalf(\Ntc-1) Y_2\tan\chi\biggr] \tpiz F_{\mu\nu}
\widetilde F^{\mu\nu}   \,,\\
\CL_{\tpiz Z\gamma} &=&
-\frac{e\sqrt{g^2+g'^2}\,(1-4\sin^2\thw)\Ntc}{16\pi^2 F_\pi} \nn\\
&& \quad \times \biggl[Y_1\cot\chi - \tthalf(\Ntc-1) Y_2\tan\chi\biggr] \tpiz
F_{\mu\nu} \widetilde Z^{\mu\nu}  \,,\\
\CL_{\tpiz ZZ} &=& \frac{(g^2+g'^2)\sin^2\thw(1-2\sin^2\thw)\Ntc}{16\pi^2
  F_\pi} \nn\\
&& \quad \times\biggl[Y_1\cot\chi - \tthalf(\Ntc-1) Y_2\tan\chi\biggr] \tpiz
Z_{\mu\nu} \widetilde Z^{\mu\nu}\,.
\eea
Recall that $N_1 = 1$ and $N_2 = 3(\Ntc-2)$ for the model in
Eq.~(\ref{eq:twoscale}) and note that $1 + 4Y_i^2 = 2(Q_{Ui}^2 + Q_{Di}^2)$,
twice the sum of the squares of technifermion $T_i$'s electric charges.
Notice also the potential for cancellations between the $T_1$ and $T_2$ terms
in these expressions that we mentioned above. This will have an especially
striking effect on $\sigma B(gg \to \etat \to \gamma\gamma)$. A similar
cancellation occurs between the $\etat \to \gamma\gamma$ and $\tpiz
\to\gamma\gamma$ amplitudes.

It is clear from these interactions that the rates for $\eta_{L,H} \to ZZ^*
\to 4\ell$ and $\eta_{L,H} \to WW^* \to \ell\nu\ell\nu$ are very much less
than $\eta_{L,H} \to \gamma\gamma$. The question of whether the $ZZ^*$ and,
to a lesser extent, the $WW^*$ data reported by ATLAS and CMS are real or
poorly understood backgrounds may be resolved by the data taken in 2012.  We
will comment on $\etal \to Z\gamma$ rate in Sec.~6. Finally, with the
complete mixing of Eq.~(\ref{eq:etapimix}), the coupling of $\eta_{L,H}$ to
two electroweak bosons $V_1$ and $V_2$ is given by
\be\label{eq:etalhVV}
\CL_{\eta_{L,H}V_1V_2} = \sqrt{\thalf}\left(\CL_{\etat V_1 V_2} \mp {\rm
    sgn}(b_2)\, \CL_{\tpiz V_1 V_2}\right)\,.
\ee

Consider the $\eta_{L,H} \bar ff$ couplings now. From
Eq.~(\ref{eq:SMcouplings}), they are determined by the ETC interactions
coupling quarks and leptons to technifermions. These are the same
interactions responsible for the SM fermions' masses (except for most of
$m_t$) and it is therefore tempting to assume that the couplings to $\bar ff$
are simply of order $m_f/\Fetat$. This is naive, however. As discussed in
Ref.~\cite{Lane:1989ej, Lane:1991qh}, a generic scenario for the fermions'
ETC couplings in a two-scale model is that SM fermions~$f$ connect to $T_1$
and $T_1$ to $T_2$. In walking technicolor, the one-loop $f$--$T_1$--$f$
graphs and the two-loop $f$--$T_1$--$T_2$--$T_1$--$f$ graphs can be
comparable. Thus, it is not at all obvious that the sum of these two
contributions to the $\etat$ and $\tpiz$ couplings to $\bar ff$ have a simple
proportionality to $m_f$. Therefore, we write
\bea\label{eq:etapiff}
\CL_{\etat \bar ff} &=& i\sum_f \frac{\zeta_{\etat,f}\, m_f}{\Fetat} \etat \bar
f\gamma_5 f\,,\nn\\
\CL_{\tpiz \bar ff} &=& i\sum_f \frac{\zeta_{\tpi,f}\, m_f}{F_\pi} \tpiz \bar
f\gamma_5 f\,,
\eea
where the factors $\zeta_f$ for $\etat$ and $\tpiz$ will have to be fixed by
experiment.\footnote{Actually, there is no reason that these Yukawa
  interactions should be parity-conserving but, for our purpose here, this
  assumption is sufficient.} % Our main concern in this paper will be with the
% $\zeta$-factors controlling $\eta_{L,H} \to \tau^+\tau^-$ and $\bar bb$.

\section*{6. $\eta_{L,H}$ Phenomenology}

We begin with a comparison of the rates of $gg$ fusion of $\eta_{L,H}$ and
the SM Higgs. The coupling of $H$ to two gluons is given to sufficient
accuracy by
\be\label{eq:Hgg}
\CL_{Hgg} = \frac{g_C^2}{48\pi^2 v} H G^\alpha_{C,\mu\nu} G_C^{\alpha,\mu\nu}\,.
\ee
Then, using $\CL_{\etat gg}$ from Eq.~(\ref{eq:etagg}), and assuming the
complete mixing of Eq.~(\ref{eq:etapimix}) and $M_{\eta_{L,H}} = M_H$, we
have
\be\label{eq:etaHratio}
\frac{\sigma(gg \to \eta_{L,H})}{\sigma(gg \to H)} =
\left(\frac{3N_1\Ntc(\Ntc-1)v}{4\sqrt{2}\Fetat}\right)^2 =
\frac{40.5}{1 + 35\sin^2\chi}\,.
\ee
The second equality is for $\Ntc = 4$, $N_1 = 1$ and $N_2 =6$. If we use the
limit $\sin\chi < 0.3$ obtained for LSTC with $M_{\trho} \simle
300\,\gev$~\cite{Collaboration:2012kk,Eichten:2012hs}, this ratio is $\simge
9.8$. This large $gg$-production rate will be compensated by a $B(\etal \to
\gamma\gamma)$ that is suppressed by the cancellation mentioned above.

\begin{figure}[!t]
 \begin{center}
\includegraphics[width=3.15in, height=3.15in]{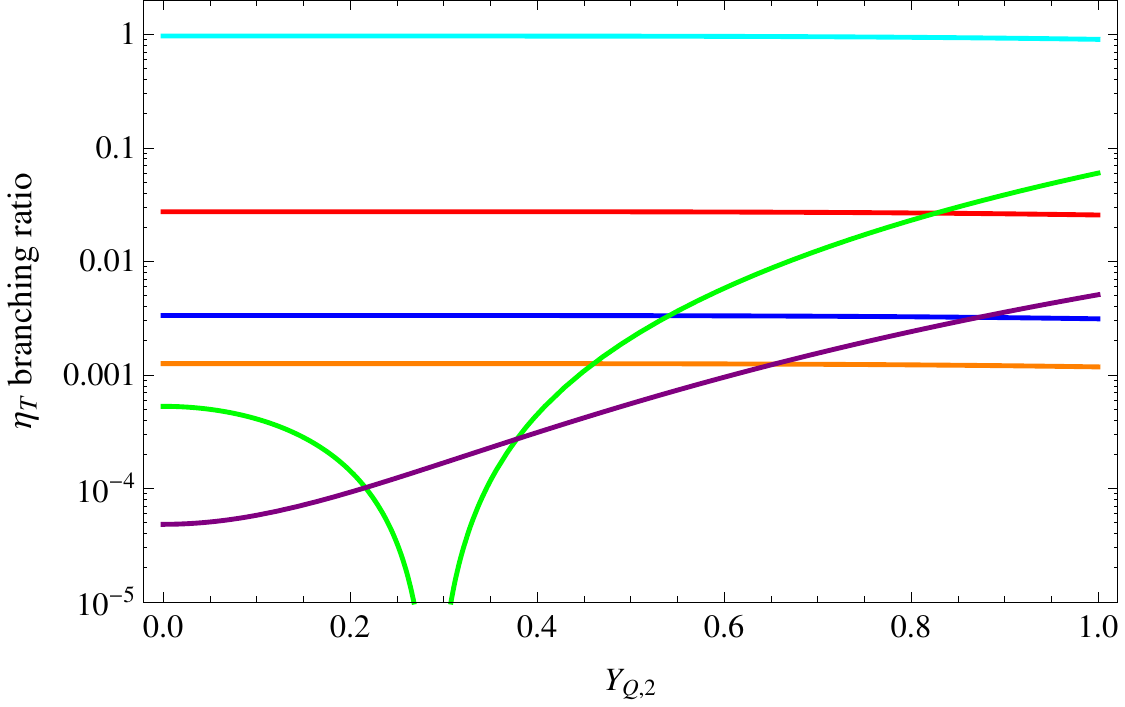}
\includegraphics[width=3.15in, height=3.15in]{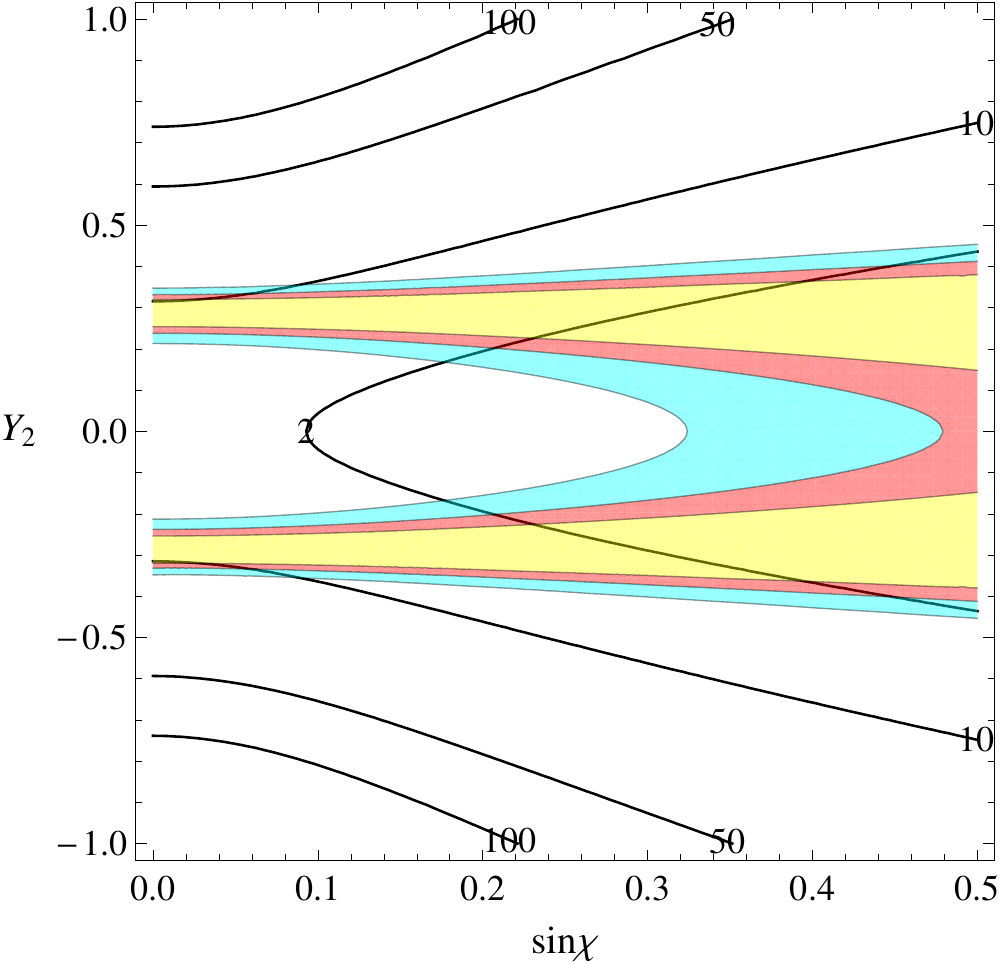}
%\vskip -0.5in
\caption{Left: The decay branching ratios as a function of $Y_2$ for a $125\,\gev$
  $\etat \to gg$ (teal), $\bar bb$ (red), $\tau^+\tau^-$ (blue), $\bar cc$
  (orange), $\gamma\gamma$ (green) and $Z\gamma$ for a real photon and
  on-shell $Z$ (purple). The $WW^*$ and $ZZ^*$ rates are negligible. See the
  text for how $\bar ff$ couplings are set. Right: The ratio $R_H = \sigma
  B(gg\to\etal\to \gamma\gamma)/\sigma B(gg\to H\to \gamma\gamma)$ for
  $M_{\etat} = M_H = 125\,\gev$, as a function of $\sin\chi$ and $Y_2$. $R_H
  < 1$ (yellow), $1.0 < R_H < 2.0$ (ochre), $2.0 < R_H < 4.0$ (teal).
  Overlaid on this plot are contours $\sigma B(gg \to \etat \to
  Z\gamma)/\sigma B(gg \to H \to Z\gamma)$.
  \label{fig:unmix}}
 \end{center}
 \end{figure}

 In the rest of this section we present results assuming both zero
 $\etat$-$\tpiz$ mixing and complete mixing. They consist mainly of the
 $\eta_{L,H}$ decay branching ratios, the ratio $\sigma B(gg\to\etal\to
 \gamma\gamma)/\sigma B(gg\to H\to \gamma\gamma)$ for $M_H = M_{\etal} =
 125\,\gev$, and $\sigma B(gg\to\etah\to \gamma\gamma)$ versus the
 $\etal$-rate. The last assumes $M_{\etah} = 180\,\gev$, a value
 corresponding to complete $\etat$-$\tpiz$ mixing at $M_{\tpipm} =
 155\,\gev$. We assume throughout that the $T_1$ hypercharge $Y_1 = 0$, which
 is strongly suggested by the absence of a signal for $\tom \ra \ellp\ellm$
 at the rate expected in LSTC for $M_{\tom} \simeq 300\,\gev$~\cite{:2012hf}.
 The value $\sin\chi = 0.3$ is used to determine $\Fetat$ and the $\tpiz$
 couplings in the branching-ratio plots; it is varied for calculating the
 branching ratios in the $\sigma B$ plots. We assume $\zeta_\tau$ and
 $\zeta_b$ factors that give the same $\sigma B$ as the SM Higgs.\footnote{In
   more detail: For a specific $\etat$-$\tpiz$ mixing, we calculate $\zeta_f$
   with $Y_1 = Y_2 = 0$. (There is only weak dependence on $Y_2$.) Solving
   $\sigma B(gg\to \etal \to \bar ff)/(\sigma B(gg \to H \to \bar ff) = 1$
   for $\zeta_\tau$ and $\zeta_b$, and taking all $\zeta_f$ equal the larger
   of the two, gives $\zeta_f$ as a function of $\sin\chi$ for each
   $\etat$-$\tpiz$ mixing. The results presented here for gauge boson
   pair-production rates (mostly diphoton) are insensitive to $\zeta_f$ so
   long as $B(\etal \to \bar ff) \simle B(H \to \bar ff)$ because the $\etat$
   width is dominated by its $gg$-decay rate.}

\begin{figure}[!t]
 \begin{center}
\includegraphics[width=3.15in, height=3.15in]{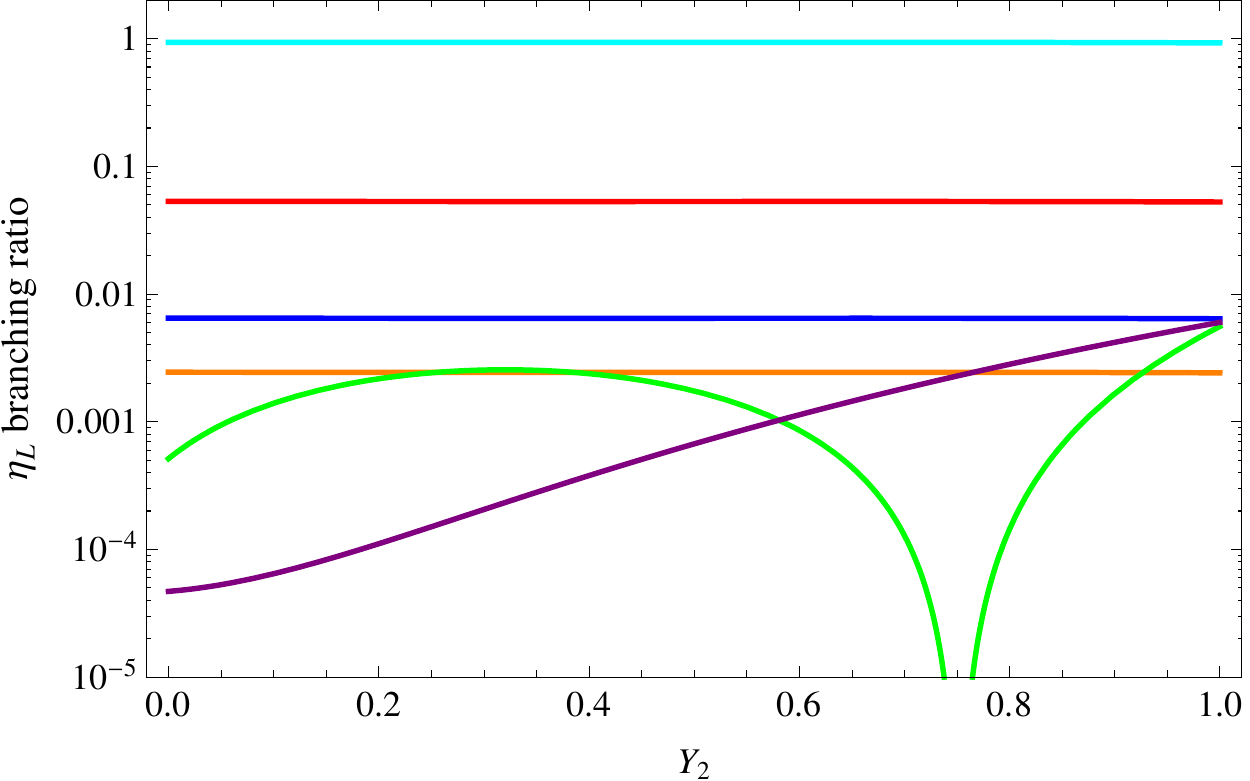}
\includegraphics[width=3.15in, height=3.15in]{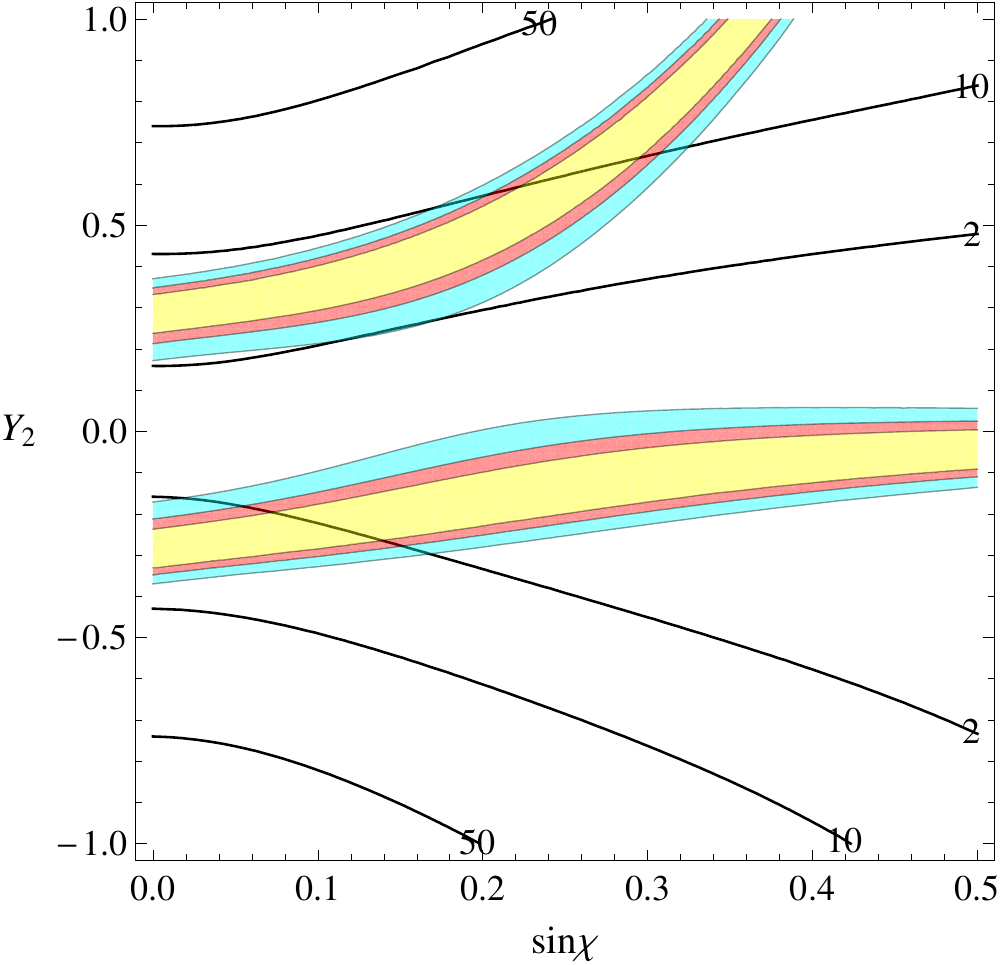}
%\vskip -0.5in
\caption{Left: The decay branching ratios as a function of $Y_2$ for a $125\,\gev$
  $\etal \to gg$ for the case of complete $\etat$-$\tpiz$ mixing with ${\rm
    sgn}(b_2) > 0$.  Right: The ratio $R_H = \sigma B(gg\to\etal\to
  \gamma\gamma)/\sigma B(gg\to H\to \gamma\gamma)$ for $M_H = M_{\etal} =
  125\,\gev$, as a function of $\sin\chi$ and $Y_2$. Overlaid on this plot
  are contours $\sigma B(gg \to \etal \to Z\gamma)/\sigma B(gg \to H \to
  Z\gamma)$. The color codes are as in Fig.~\ref{fig:unmix}.
  \label{fig:posmix}}
 \end{center}
 \end{figure}

 The $\etat$ branching ratios and $\gamma\gamma$ production rate are shown in
 Fig.~\ref{fig:unmix} for the case of no mixing with $\tpiz$. For $Y_1 = 0$,
 these are even functions of $Y_2$. As anticipated, the zero in the
 $\gamma\gamma$ rate at $Y_2 = 0.29$ is due to a cancellation between the
 $T_1$ and $T_2$ contributions. For $\sin\chi < 0.3$, there are narrow bands
 ($\Delta Y_2 \simeq 0.07$) centered on $Y_2 = \pm 0.29$ where the $\etat \to
 \gamma\gamma$ rate is up to four times as large as the SM Higgs rate. We
 expect that any additional jets associated with this $gg$-production would
 be color-connected with the primary production and not exhibit a rapidity
 gap. To our knowledge, there is no published analysis of this. Contours
 giving the ratio $\sigma B(gg \to \etat \to Z\gamma)/\sigma B(gg \to H \to
 Z\gamma)$ are overlaid on this plot. The ratio is 2--10 for $\sin\chi <
 0.3$. We have estimated the rate for $\etat \to Z\gamma^* \to 4\ell$ and
 found that, for a luminosity of $10\,\ifb$, at most half an event would have
 been produced. After efficiencies, essentially none of the events in
 Figs.~\ref{fig:CMSMZ1vMZ2},\,\ref{fig:ATLASMZ1vMZ2} could be due to this
 $\etat$ decay.

%%  I have the following calculation: take eta -> Zgamma^* -> Z l+l- for
%%  arbitrary eta Z gamma coupling, calculated as three-body decay and imposing
%%  a limit of 5 gev on m_ll.  plugging in the eta Z gamma coupling in the model
%%  and dividing by the width, I have the BR(eta -> Z gamma -> Zll). multiplying
%%  by luminosity, cross section, etc. I get the number of Z + l l events at the
%%  right luminosity. Adding in BR(Z-> ll) and factors of 2 for e and mu gets me
%%  the number, approximately, of 4 lepton events. For 5 fbinv, this gets me
%% ~0.2 events in the parameter range we care about .. adding in a more
%% realistic analysis and efficiency numbers would lower this further, probably
%% another factor of 2 or 3.
 
 The $\etal$ branching ratios and $\gamma\gamma$ rate compared to the SM
 Higgs are shown for the complete-mixing cases and $Y_1 = 0$ in
 Fig.~\ref{fig:posmix} for ${\rm sgn}(b_2) > 0$ and Fig.~\ref{fig:negmix} for
 ${\rm sgn}(b_2) < 0$. These two cases go into each other by reversing the
 signs of $Y_1$ and $Y_2$. For $Y_1 = 0$ and $\sin\chi = 0.3$, the zero in
 $B(\etal \to \gamma\gamma)$ for $Y_2 > 0$ occurs for the two cases at $0.75$
 and $0.11$, respectively. Note that $\etal$ decay rates are dominated by
 those for $\etat$, in particular $B(\etal \to gg) \simeq 100\% \gg B(\etat
 \to \bar ff)$. Still, as explained in footnote~10, we have chosen $\etal$
 couplings to fermions so that $\sigma(gg \to \etal \to \bar bb\,\,{\rm
   or}\,\, \tau^+\tau^-)/\sigma(gg \to H \to \bar bb\,\,{\rm
   or}\,\,\tau^+\tau^-) \sim 1$. The allowed ranges of $\sigma B(\etal \to
 \gamma\gamma)$ occur in bands of thickness $\Delta Y_2 \simeq 0.2$ and, for
 the two mixing cases, they are mirror reflections of each other about $Y_2 =
 0$ for $Y_1 = 0$. In these allowed regions, $B(\etal \ra \gamma\gamma)$ is
 4--10 times smaller than the SM Higgs branching ratio. As in the unmixed
 case, the $\etal \to Z\gamma$ rate is 2--10 times the SM Higgs rate, much
 too small to account for the data in
 Figs.~\ref{fig:CMSMZ1vMZ2},\ref{fig:ATLASMZ1vMZ2}.

\begin{figure}[!t]
 \begin{center}
\includegraphics[width=3.15in, height=3.15in]{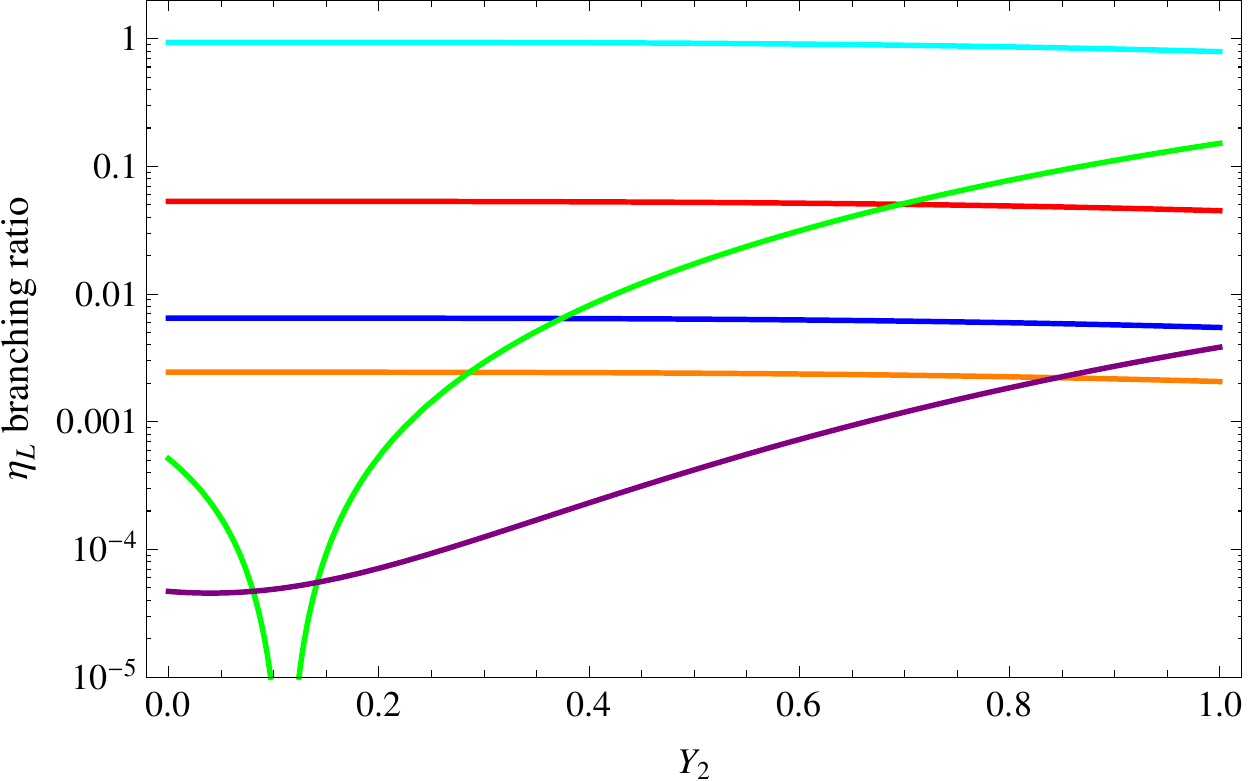}
\includegraphics[width=3.15in, height=3.15in]{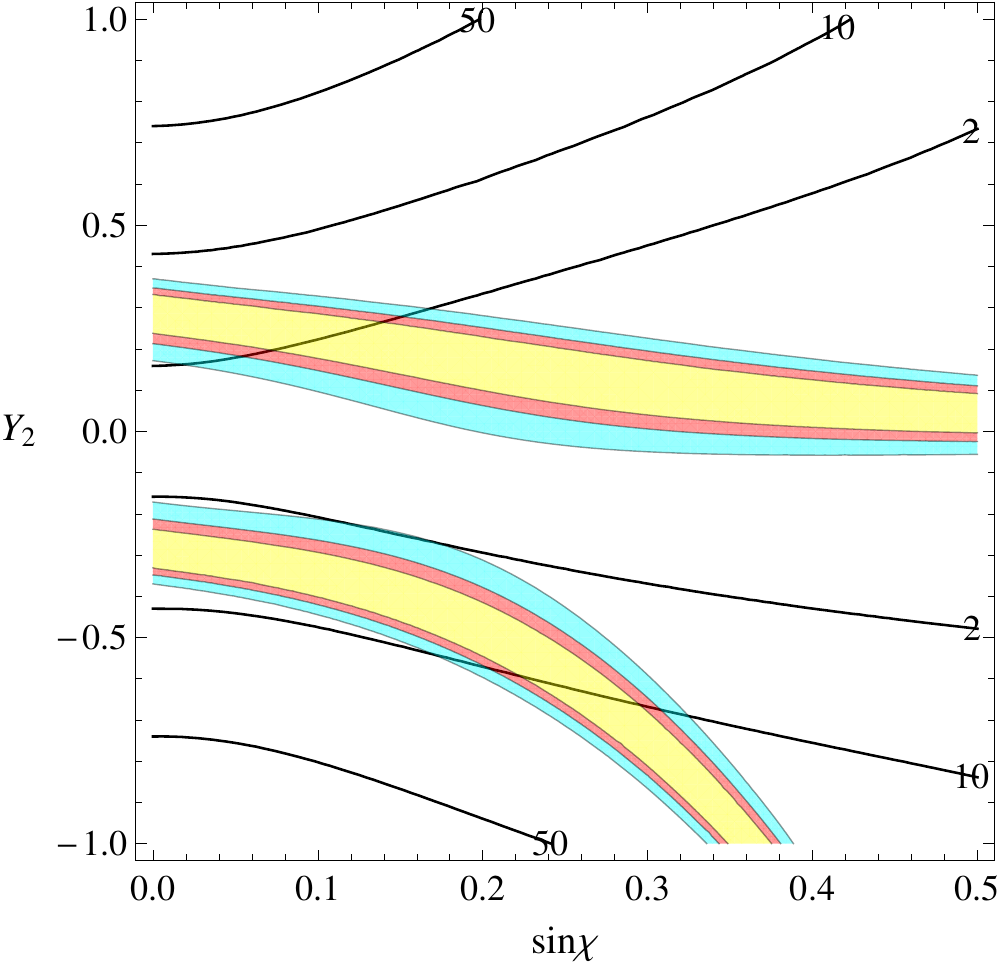}
%\vskip -0.5in
\caption{Left: The decay branching ratios as a function of $Y_2$ for a $125\,\gev$
  $\etal \to gg$ for the case of complete $\etat$-$\tpiz$ mixing with ${\rm
    sgn}(b_2) < 0$.  Right: The ratio $R_H = \sigma B(gg\to\etal\to
  \gamma\gamma)/\sigma B(gg\to H\to \gamma\gamma)$ for $M_H = M_{\etal} =
  125\,\gev$, as a function of $\sin\chi$ and $Y_2$. Overlaid on this plot
  are contours $\sigma B(gg \to \etal \to Z\gamma)/\sigma B(gg \to H \to
  Z\gamma)$. The color codes are as in Fig.~\ref{fig:unmix}.
  \label{fig:negmix}}
 \end{center}
 \end{figure}
\begin{figure}[!t]
 \begin{center}
\includegraphics[width=3.15in, height=3.15in]{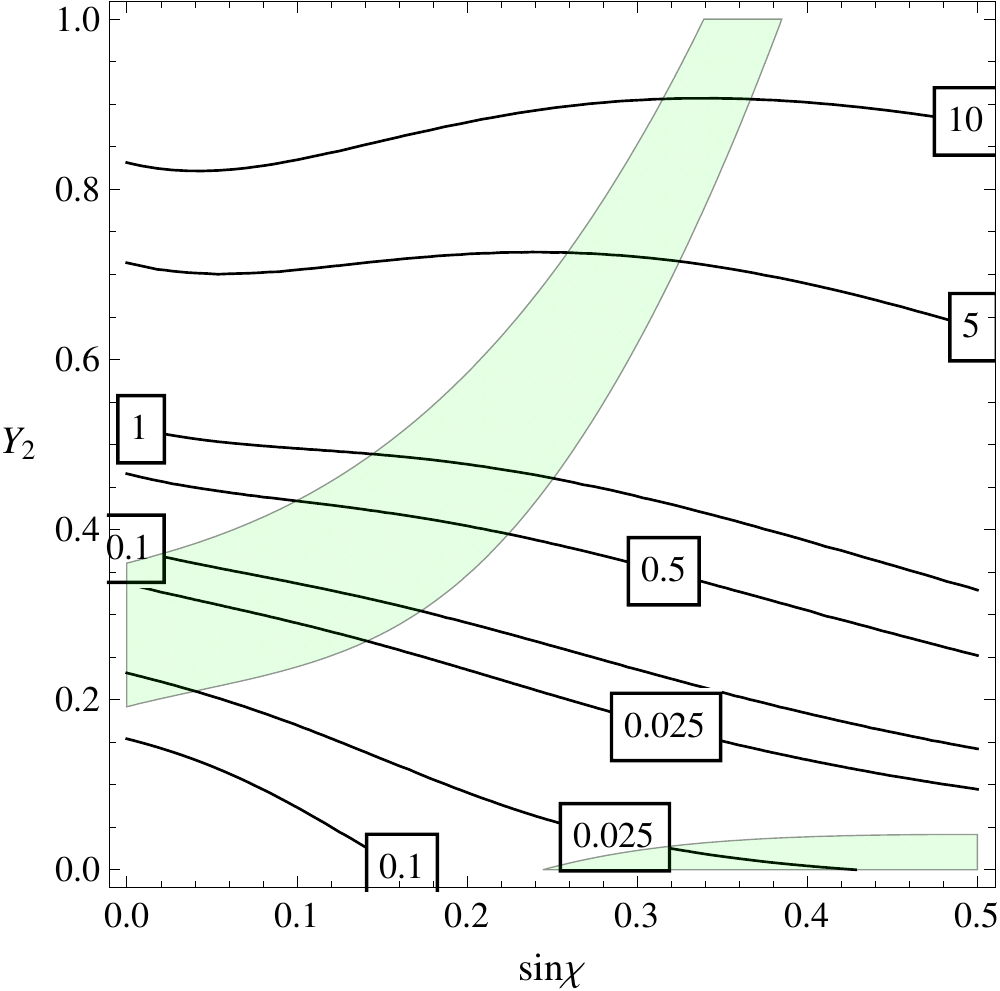}
\includegraphics[width=3.15in, height=3.15in]{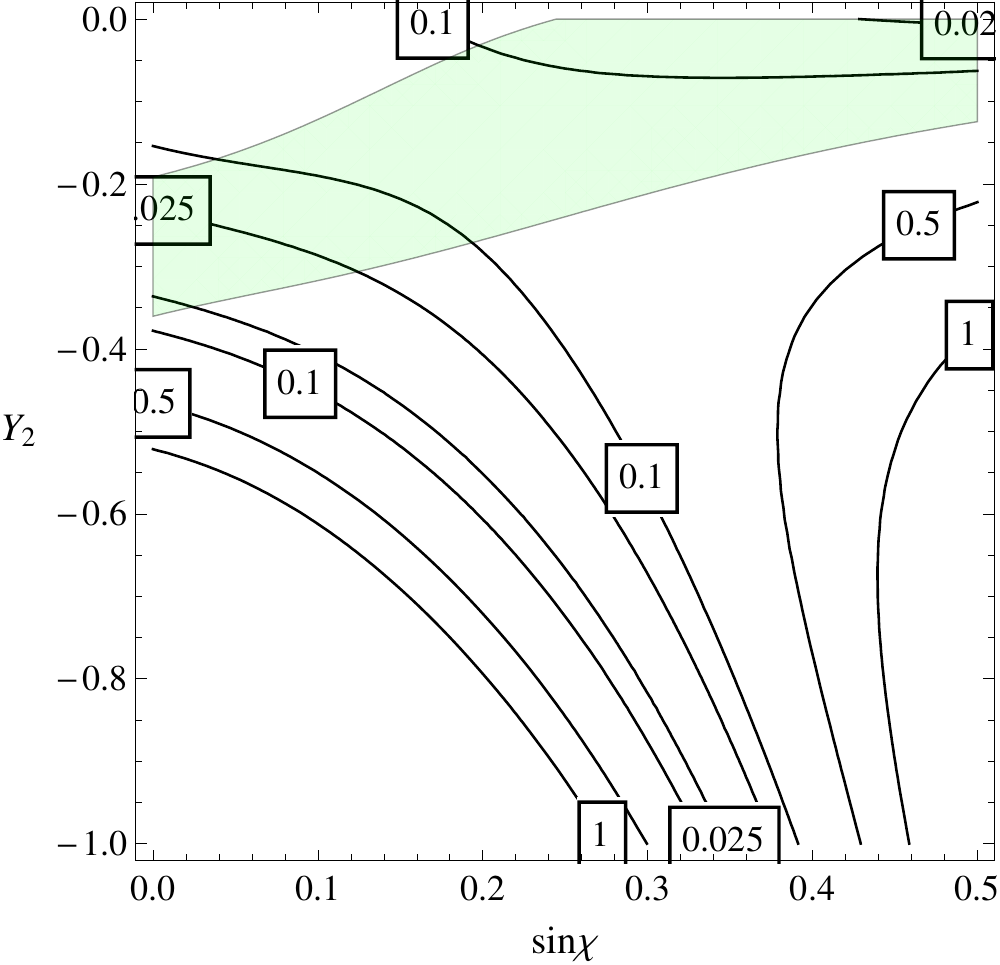}
%\vskip -0.5in
\caption{The green-shaded regions are $R_H = \sigma B(gg\to\etal\to
  \gamma\gamma)/\sigma B(gg\to H\to \gamma\gamma) < 4$ for $M_H = M_{\etal} =
  125\,\gev$ and ${\rm sgn}(b_2) > 0$, as a function of $\sin\chi$ and
  $Y_2$. Overlaid on these plots are contours $\sigma B(gg \to \etah \to
  \gamma\gamma)$, in picobarns, for $M_{\etah} = 180\gev$.
  \label{fig:sigLvsigH}}
 \end{center}
 \end{figure}

 Finally, in Fig.~\ref{fig:sigLvsigH} we overlay the $\sigma B(gg \to \etal
 \to \gamma\gamma)/\sigma B(gg \to H \to \gamma\gamma)$ ratios with contours
 of $\sigma B(gg \to \etah \to \gamma\gamma)$ given in picobarns. Based on a
 CMS search for diphoton resonances in $2.2\,\ifb$ of 7-TeV
 data~\cite{Chatrchyan:2011fq}, we estimate that $\sigma B(gg \to \etah \to
 \gamma\gamma) \simle 0.25\,\pb$ is allowed. This is consistent with both
 branches of the green-shaded region of this figure for $|Y_2| < 0.4$.

%%  I took the Madgraph implementation of the eta_T and generated events for
%%  Meta = 180 gev. Applying the cuts that CMS used in their 1112.0688
%%  anaylsis
%% Ref.\cite{Chatrchyan:2011fq}
%%  I can calculate the accpetance & use their quoted efficiency. CMS quotes the
%%  events in a M_gamma gamma bin from 140 -200 gev. Using the expected and
%%  observed number of events and the uncertainty, I get a limit of about 150
%%  events allowed in that Mgaga bin at 2.2 fbinv. Using the accpetance&
%%  efficiencies from my mock analysis, this corresponds to an allowed cross
%%  section of about 250 fb. This is probably on the conservative side as I
%%  didnt account for smearing or detector inefficiencies, but I would not go
%%  above 500 fb.

\section*{7. $\etat$-$\tpiz$ Mixing and LSTC Collider Phenomenology}

%% If $\etat$-$\tpiz$ mixing is large, the LSTC description of CDF's dijet
%% excess is greatly modified.
The discussion in this section is based on our interpretation of CDF's dijet
excess as the production of a 280--$290\,\gev$ $\tro$ which decays to a
150--$160\,\gev$ $\tpi$ plus a $W$-boson~\cite{Aaltonen:2011mk,CDFnew,
  Eichten:2011sh, Eichten:2012hs}.\footnote{In Ref.~\cite{Eichten:2011sh} we
  found that $\simeq 25\%$ of the Tevatron signal was due to $\ta \to W\tpi$,
  with $M_{\ta} = 1.1 M_{\tro}$ assumed.} The $\tpi$ decays 90--95\% of the
time to $\bar qq$ jets (which may or may not contain $b$-jets, hence the
spread we assume in $M_{\tpi}$ and $M_{\tro}$). With large $\etat$-$\tpiz$
mixing, the $\tropm \to W\tpiz$ component of the $150\,\gev$ dijet signal is
absent. To some extent, this loss is replaced by $\tro \ra W\etal \to
\ellpm\nu jj$ with $\Mjj \simeq 125 \,\gev$. Since this decay is dominated by
its $\tro \to W\tpiz$ component, the dijets are mainly $\bar qq$ jets.
Detailed calculation of this new phenomenology requires either a complete
rewrite of the {\sc Pythia} code for LSTC or a new implementation in another
amplitude generator because changes in the $\tro$ partial widths make it
difficult to guess individual production rates. This is beyond the scope of
this paper. Here we will be satisfied with a list of the important changes we
anticipate.

\begin{itemize}
  
\item[1)] The rate for $\tro,\ta \to W\tpi$ is likely to be reduced. Simply
  (and naively) eliminating the $W\tpiz$ mode results in about a 35\%
  reduction of the dijet excess signal~\cite{Eichten:2011sh}.
  
\item[2)] There will be a $\tro,\ta \to W\etal \to \ell\nu jj$ signal at
  $\Mjj \simeq 125\,\gev$, largely due to its $W\tpiz$ component. The dijet
  peak may overlap somewhat with the $\troz \to W^\pm\tpimp$ dijet excess.
  While $\tro,\ta \to W\etal$ is suppressed by the mixing, it is enhanced by
  the greater phase space and, so, may not be much smaller that the
  $W^\pm\tpimp$ rate. Note that this will appear as associated production of
  $\etal$ with $W$, but the $Wjj$ invariant mass will peak near $M_{\tro}$.
  There is {\em no} significant associated production of $\etal$ with $Z$.
  
\item[3)] The channel $\tropm \to \tpipm\etal$ is open and the $\tpiz$
  component of this amplitude is a strong process, unsuppressed by
  $\sin\chi$.  Even though the $Q$-value for this decay is only $\sim
  5\,\gev$, this mode could be an important part of the $\tropm$ width and
  its production rate might be as large $\sim 500\,\fb$ at the LHC. We do not
  know of any limit on this four-jet process, especially since the two dijets
  have rather different masses.
  
\item[4)] A primary signal at the LHC for confirming the CDF dijet excess is
  $\tropm \to Z\tpipm \to \ellp\ellm jj$. In Ref.~\cite{Eichten:2012hs}, we
  predicted a rate of $190\,\fb$ for this final state ($220\,\fb$ for $Y_1 =
  0$ and $\sin\chi = 0.3$). This rate is likely reduced by the open
  $\tpipm\etal$ channel; a {\em rough} estimate is a 50--60\% reduction. This
  is an unfortunate hit to an otherwise very promising channel for the 2012
  data.
  
\item[5)] A similar reduction in the rate for $\tropm \to WZ \to 3\ell\nu$ or
  $\ellp\ellm jj$ is to be expected. This would weaken the bound $\sin\chi <
  0.3$ implied by the recent CMS data~\cite{Collaboration:2012kk}. On the
  other hand, the idea of low-scale TC does not make much sense if $\sin\chi
  \simge \half$.
  
\item[6)] Last, though not least, we again urge a search for $\etah \to
  \gamma\gamma$ near $180\,\gev$. Over most of the allowed regions in
  Fig.~\ref{fig:sigLvsigH}, $\sigma B(\etah \to \gamma\gamma) \simle
  0.25\,\pb$ at the LHC. The upper end of this range should be accessible
  soon---if not already excluded.

\end{itemize}

\section*{8. Conclusions}

The ``Higgs impostor'' proposal made in this paper is motivated both by our
desire for a technicolor explanation for the new boson $X(125)$ and by the
apparent differences between the ATLAS and/or CMS data and what is expected
for a Higgs boson. The most important discrepancy is the $ZZ^* \to 4\ell$
data of both experiments, a channel valued for its high mass resolution. The
low number of what might be called ``gold-plated'' events in the CMS data ---
those which appear to contain a real, on-shell $Z$-boson and which fall in
the dark ``signal region'' of the three distributions, $M_{Z1}$ and $M_{Z2}$
versus each other and $M_{4\ell}$ --- is one glaring example. The ATLAS $ZZ^*
\to 4\ell$ data appears to have a similar deficit of gold-plated events. A
second example is the rather large fluctuations in the number of events in
the signal region of $M_{Z1}$ vs.~$M_{Z2}$ between the July and
November/December~2012 data releases by both experiments. All this may just
be statistics at work and be resolved in favor of the popular Higgs
description when the next large batch of data is released. But, as we said at
the outset, the SM Higgs outcome would confront theorists anew with the
thorny questions of naturalness, hierarchy and flavor. If, on the other hand,
the discrepancies in the data are real, then we may, at long last, have begun
to unravel the mystery of electroweak symmetry breaking. That is a lot to
hope for.

In this paper we proposed an alternative to the SM Higgs interpretation:
$X(125)$ is a technipion, $\etal$. Our proposal has several immediately
testable consequences in addition to discrediting the $X\to ZZ^*,WW^*$
data. Chief among these is that there is likely to be another Higgs impostor
state $\etah$ which is not far from 200~GeV and which may be visible in the
diphoton spectrum. If the CDF dijet excess is real, and our LSTC
interpretation of it correct, then $M_{\etah} =
170$--$190\,\gev$. Furthermore, the $\Mjj$ spectrum in the range $\sim
100$--$150\,\gev$ range is contaminated by a sizable $\tro \to W\etal$
component that will complicate modeling in terms of standard diboson
production as done, e.g., by CMS in Ref~\cite{:2012he}. Finally, if all this
is correct, we expect the LSTC phenomenology presented in
Ref.~\cite{Eichten:2012hs} to be modified substantially.

\section*{Acknowledgments} 

We are grateful to T.~Appelquist, W.~Bardeen, K.~Black, T.~Bose,
P.~Catastini, A.~DeRoeck, C.~Fantasia, C.~Hill, K.~Terashi, B.~Zhou and J.~Zhu
for valuable conversations and advice. This work was supported by Fermilab
operated by Fermi Research Alliance, LLC, U.S.~Department of Energy
Contract~DE-AC02-07CH11359 (EE and AM) and in part by the U.S.~Department of
Energy under Grant~DE-FG02-91ER40676~(KL). KL's research was also supported
in part by Laboratoire d'Annecy-le-Vieux de Physique Theorique (LAPTh) and
the CERN Theory Group and he thanks LAPTh and CERN for their hospitality.

\vfil\eject

\bibliography{eta_T_2}
\bibliographystyle{utcaps}
\end{document}